\documentclass[]{emulateapj}
\usepackage{natbib,graphicx,psfig}

\def\boldxi{\mbox{\boldmath $\xi$}}
\def\boldnabla{\mbox{\boldmath $\nabla$}}

\def\boldOmega{\mbox{\boldmath $\Omega$}}
\def\bnabla{\mbox{\boldmath $\nabla$}}

\def\Bruntfreq{Brunt-V{\"a}is{\"a}l{\"a}\,\,\,}
\def\refnew#1{(\ref{#1})}
\def\be{\begin{equation}}
\def\ee{\end{equation}}
\def\ba{\begin{eqnarray}}
\def\ea{\end{eqnarray}}

\def\erg{\, \rm erg}
\def\K{\, \rm K}
\def\s{\, \rm s}
\def\km{\, \rm km}
\def\cm{\, \rm cm}
\def\g{\, \rm g}
\def\m{\, \rm m}
\def\dyne{\, \rm dyne}

\def\pomega{\varpi}

\begin{document} 
	
\title{\mbox{Origin of Tidal Dissipation in Jupiter:
II. the Value of $Q$}}

\lefthead{Tidal $Q$ of Jupiter}
\righthead{Wu}
\righthead{}

\author{Yanqin Wu}
\affil{Department of Astronomy and Astrophysics, 60 St. George Street, 
	University of Toronto, Toronto, ON M5S 3H8, Canada}

\begin{abstract}
The process of tidal dissipation inside Jupiter is not yet understood.
Its tidal quality factor ($Q$) is inferred to lie between $10^5$ and
$10^6$.  Having studied the structure and properties of inertial-modes
in a neutrally buoyant, core-less, uniformly rotating sphere
\citep{2004a}, we examine here their effects on tidal dissipation. The
rate of dissipation caused by resonantly excited inertial-modes
depends on the following three parameters: how well they are coupled
to the tidal potential, how strongly they are dissipated (by the
turbulent viscosity), and how densely distributed they are in
frequency. We find that as a function of tidal frequency, the $Q$
value exhibits large fluctuations, with its maximum value set by the
group of inertial-modes that satisfy $\delta \omega \sim \gamma$,
where $\delta\omega$ is the group's typical offset from an exact
resonance, and $\gamma$ their turbulent damping rates.
These are intermediate order inertial-modes with wave-number $\lambda
\sim 60$ and they are excited to a small surface displacement amplitude of
order $ 10^3 \cm$. The $Q$ value drops much below the maximum value
whenever a lower order mode happens to be in resonance.  In our model,
inertial-modes shed their tidally acquired energy very close to the
surface within a narrow latitudinal zone (the 'singularity belt'), and
the tidal luminosity escapes freely out of the planet.

Strength of coupling between the tidal potential and inertial-modes is
sensitive to the presence of density discontinuities inside Jupiter.
In the case of a discreet density jump, as may be caused by the
transition between metallic and molecular hydrogen, we find a
time-averaged $Q \sim 10^7$, with a small but non-negligible chance
($\sim 10\%$) that the current $Q$ value falls within the empirically
determined range. Whereas when such a jump does not exist, $Q \sim
10^9$. Even though it remains unclear whether tidal dissipation due to
resonant inertial-modes is the correct answer to the problem, it is
impressive that our simple treatment here already leads to three to
five orders of magnitude stronger damping than that from the
equilibrium tide.

Moreover, our conclusions are not affected by the presence of a small
solid core, a different prescription for the turbulent viscosity, or
nonlinear mode coupling, but they depend critically on the static
stability in the upper atmosphere of Jupiter. This is currently
uncertain. Lastly, we compare our results with those from a competing
work by \citet{gordon} and discuss the prospect of extending this
theory to exo-jupiters, which appear to possess $Q$ values similar to
that of Jupiter.
\end{abstract}


\setcounter{equation}{0}

\section{INTRODUCTION}
\label{sec:nl-intro}

\subsection{the Puzzle}
\label{subsec:observation}

We tackle the classical problem of tidal dissipation in Jupiter.  In
the following, we briefly review the problem, both for Jupiter and
for close-in extra-solar planets. For a contemporary and expansive
overview of this issue, including a detailed discussion of previous
work, we refer the readers to \citet[][hereafter {\bf OL}]{gordon}.

As Jupiter spins faster than the orbital motion of its nearest
satellite (Io), Io raises a time-dependent tide on Jupiter. The
dissipation of this tide in Jupiter transfers its angular momentum to
Io and spins down Jupiter. We adopt the convention of quantifying the
inefficiency of the dissipation\footnote{This assumes $Q$ is
independent of the orbital phase.  \citet{hut} and others have adopted
instead a constant lag time $\tau = \epsilon/(\Omega - {\dot f})$,
where $\Omega$ and $\dot f$ are the rotational and instantaneous
orbital angular velocity, respectively, and $f$ is the free
anomaly. These two approaches are comparable if $Q$ is frequency
independent and if the eccentricity is not too large.} by a
dimensionless quality factor $Q$, which is the ratio between the
energy in the (equilibrium) tide ($E_0$, see \S
\ref{subsubsec:equitide}) and the energy dissipated per period
\be
Q \equiv {2\pi E_0 \over{\oint - {{dE}\over{dt}} dt}} = {1\over{\tan 2
\epsilon}} \approx {1\over 2\epsilon},
\label{eq:defineQ}
\ee
where $\epsilon$ is called the lag angle. It corresponds to the angle
between the directions of Io and the tidal bulge when we are concerned
with the equilibrium tide.
The rate of tidal synchronization scales inversely linearly with $Q$ \citep{murray},
\be
{{d \Omega}\over{dt}} = - {\mbox{SIGN}} (\Omega - \omega_{\rm Io}) {{3
k_2} \over{2 \alpha Q}} \left({{M_{\rm Io}}\over{M_J}}\right)^2
\left({{R_J}\over a}\right)^3 \omega_{\rm Io}^2,
\label{eq:synchro}
\ee
where $\Omega$ is Jupiter's spin frequency, $\omega_{\rm Io} = (G
M_{\rm J}/a^3)^{1/2}$ is Io's orbital frequency, $a$ its orbital
separation, $M_J$, $R_J$, $k_2$, $\alpha$ are Jupiter's mass, radius,
tidal love number ($k_2 \approx 0.38$) and moment of inertia constant
($\alpha = I/M_J R_J^2\sim 0.25$), respectively. $M_{\rm Io}$ is Io's
mass.

Based on the current resonant configuration of the Galilean
satellites, Jupiter's $Q$ value has been estimated to be $10^5 \leq Q
\leq 2\times 10^6$ with the actual value likely closer to the lower
limit \citep{goldreichsoter,peale}. The interior of Jupiter is
comprised of (at most) a small heavy-element core, a metallic hydrogen
region and a molecular hydrogen envelope \citep[see,
e.g.][]{guillotreview}, with convection being the dominant heat
transfer mechanism outside the core.  The most reliable theoretical
estimate for the $Q$ value -- based on turbulent viscosity acting on
the equilibrium tide -- puts $Q \sim 10^{13}$ \citep{goldreichphil},
well above the inferred value. The physical origin for this low $Q$
value (and thus higher than expected dissipation) has remained elusive
for a few decades, with suggestions ranging from a substantial inner
core \citep{dermott}, to helium hysteresis around the depth of
hydrogen metallic phase transition
\citep{stevenson}, to a postulated stratification in the interior that
harbors rotationally-modified gravity-modes
\citep{ioannou}.
Each proposal promises interesting implication for the physics of
dense matter or for the structure of Jupiter. Where does the truth
lie? Intriguingly, Saturn's inferred $Q$ value is similar to that of Jupiter
\citep{goldreichsoter}.

The discovery of close-in extra-solar jupiters has rejuvenated our
interest in this problem and provided new insights. While the majority
of exo-planets are in eccentric orbits around their host stars, the
closest-in ones have low or nearly zero eccentricities. This results
from the dissipation of stellar tide inside the planets which converts
orbital energy into heat without removing orbital angular
momentum. Orbital circularization due to tidal dissipation inside the planet
proceeds at a rate \citep{hut}:
\be
{1\over e}{{de}\over{dt}} = - {{27
k_2} \over{2 Q}} \left({{M_{*}}\over{M_J}}\right)
\left({{R_J}\over a}\right)^5 \omega,\label{eq:circular}
\ee
where $k_2$, $Q$ and $R_J$ are the planet's tidal love number, tidal
quality factor and radius, respectively. It orbits its host star (mass
$M_*$) with a semi-major axis $a$ and an orbital frequency $\omega$.
Fig. 1 in \citet{wu03} shows that the observed upper envelope of
planet eccentricity as a function of its semi-major axis can be
explained by a tidal quality factor of $Q \approx 3\times 10^5$ if
these are gaseous planets similar to Jupiter in their ages and
sizes.\footnote{One exception is the planet HD $80606b$ whose
abnormally high eccentricity may be acquired relatively recently
\citep{wumurray}.} 

The close-in exo-planets and Jupiter may well have different formation
history, leading to different core sizes and different interior
compositions. They certainly evolve in very different thermal
environments, resulting in diverging thermal structure in their upper
atmosphere. Nevertheless, they share similar $Q$ factors.  This
prompts us to seek a physical explanation for $Q$ which is based on
overt similarities between these planets. The first trait in common
which we believe is important is that their interiors are fully
convective. The second trait is that they rotate fast. Jupiter spins
roughly four times for every Io orbit, while the spin of close-in ($a
< 0.1 $AU) planets should have long been (pseudo-)synchronized with
their orbital motion. So in both cases, the (dominant) tidal forcing
frequencies viewed in the planets' rotating frame are below $2
\Omega$.\footnote{In this respect, it is interesting to point out that
tidally circularizing solar-type binaries have convective envelopes
and likely spin fast. Curiously, they exhibit similar $Q$ values as
these giant planets \citet{mathieu}.}  Could these two common traits
be responsible for the tidal $Q$ values?

\subsection{The Inertial-Mode Approach}
\label{subsec:past}

In a spinning and neutrally buoyant fluid sphere, a new branch of
eigen-modes arise: the inertial-modes. Their motion is restored not by
pressure or buoyancy, but by Coriolis force.  In the rotating frame,
these modes have frequencies ranging from zero to twice the spin
frequency. As noted above, the tidal frequencies also fall in this
range. How does the presence of these modes affect tidal dissipation?

We have previously studied inertial-modes in non-uniform density
spheres \citep[][hereafter Paper I]{2004a}, focusing on properties
relevant to tidal dissipation. 
We found that inertial-modes which can couple to the tidal potential
are much denser in frequency space compared to gravity- or
pressure-modes, allowing for good resonance with the tidal
forcing. Inertial-modes have unique ``singularity belts'' near the
surface where both mode amplitudes and velocity shear are the largest,
leading to strong turbulent dissipation.
Both these facts suggest that inertial-modes are good candidates to
explain the tidal dissipation in planets. In this paper, we explore
this possibility for Jupiter.

Because of mathematical difficulties, rotation has been largely
ignored in tidal theories \citep[for an exception, see][as well as
their subsequent papers]{savonije}. However, this can not be justified
when rotational frequency is comparable to or faster than the tidal
frequency. Tidal response of the fluid is strongly influenced by
rotation. Our results here show that when rotation is taken into
account, even the most rudimentary treatment gives orders of magnitude
stronger tidal dissipation than when it is not.

In this direction, most noteworthy is a recent independent work by OL,
which appeared while we were writing up our results.  In this paper,
OL calculated the effect of inertial-modes in planets, based on
essentially the same physical picture as we consider here.  We discuss
their work in the context of our results. For un-initiated readers, we
recommend their excellent and helpful review for issues related to
tidal dissipation and to inertial-modes.

\subsection{Organization}
\label{subsec:thiswork}

Paper I has laid a foundation by studying properties of
inertial-modes. In \S \ref{sec:inertial}, we first summarize results
from that paper, then proceed to discuss two issues of importance,
i.e., how strongly an inertial-mode is coupled to the tidal potential,
and how strongly an inertial-mode is damped by turbulent
viscosity. Relevant contents of a highly technical nature are
presented in Appendix
\ref{sec:polyn2}, where a simple toy model is constructed to help
explain the results.  In \S \ref{sec:Qvalue}, we discuss the effects
of inertial-mode dynamical tide on the tidal $Q$ factor, using
equilibrium tide as a comparison to illustrate the advantage of
inertial-modes.  Lastly, we discuss uncertainties in our model,
and compare our results with previous work (\S
\ref{sec:discussion}). We summarize and discuss other possible applications 
in \S \ref{sec:summary}.


\section{Inertial-Modes -- Relevant Properties}
\label{sec:inertial}

In Paper I, we show that by introducing the ellipsoidal coordinates
\citep{bryan}, the partial differential equation governing
inertial-modes can be separated into two ordinary differential
equations, both when the density is uniform \citep{bryan}, and when
the density satisfies a power-law ($\rho \propto [1-(r/R)^2]^\beta$,
where $R$ is the planet radius and $r$ the spherical
radius). Moreover, for spheres of smooth but arbitrary density laws,
we find that one could obtain sufficiently accurate (albeit
approximate) eigenfunctions using these coordinates.

Each inertial-mode in a sphere can then be characterized by three
quantum numbers, $n_1 , n_2$ and $m$. Here, $n_1$ and $n_2$ are the
number of nodes along the $x_1$ or $x_2$ ellipsoidal coordinate lines,
and $m$ is the usual azimuthal number. All perturbations satisfy the
form $e^{im \phi}$ with $\phi$ being the azimuthal angle.  For a
graphical presentation of an inertial-mode, see Figs. 4 \& 5 in Paper
I. We also introduce in Paper I the dimensionless mode wavenumber
$\lambda \sim 2(n_1 + n_2)$, which is related to the dimensionless
mode frequency $\mu = \omega/2\Omega \approx \sin (n_1 \pi/\lambda)$,
where $\omega$ is the inertial-mode frequency viewed in the rotating
frame, $\Omega$ the spin frequency, and $0 < \mu \leq 1$. Under this
convention, $m < 0$ denotes retrograde modes, while $m > 0$ prograde
ones.

\subsection{Goodness of Resonance}
\label{subsec:frequency}

In a non-rotating star, each eigenmode is identified by three quantum
numbers $n,\ell,m$ where $n$ is the number of nodes in the radial
direction, and $\ell, m$ relate to a single spherical harmonic
function $P_\ell^m(\theta,\phi)$ that describes the angular dependence
of the mode.  In contrast, the angular dependence of each
inertial-mode is composed of a series of such spherical harmonic
functions. This has the consequence that while only the $\ell = 2$,
$m=-2$ branch of non-rotating modes can be driven by a potential
forcing of the form $P_2^{-2}$ (the dominant tidal forcing term), {\bf
every} even-parity inertial-mode can potentially be driven. In this
sense, the frequency spectrum of inertial-modes is dense, and the
probability of finding a good frequency match (mode frequency
$\approx$ forcing frequency) is much improved over the non-rotating
case.

For a given forcing frequency $\mu_0$, how far in frequency does the
closest inertial-mode lie? We limit ourselves to inertial-modes with
$\lambda \sim 2 (n_1 + n_2)
\leq \lambda_{\rm max}$.  Approximate mode frequency by $\mu \approx
\sin(n_1\pi/\lambda) \sim n_1\pi/\lambda$. Modes with the same $n_2$
but different $n_1$ are spaced in frequency by $\sim \pi/\lambda \geq
\pi/\lambda_{\rm max}$. 
Now allow $n_2$ to vary between $1$ and $\sim \lambda_{\rm
max}/4$,\footnote{Unless $\mu \sim 0$ or $\mu \sim 1$, we have $n_1
\sim n_2$.} we find that the best frequency off-resonance to $\mu_0$ is
typically
\be
(\delta\mu)_{\rm min} = {{\delta \omega}_{\rm min}\over{2 \Omega}}
\sim {{\pi}\over{\lambda_{\rm max} n_2}} 
\sim {{4\pi}\over{\lambda_{\rm max}^2}}.
\label{eq:bestdmu}
\ee
For comparison, gravity- or pressure-modes in non-rotating bodies can
at best have a frequency detuning of $\delta \omega/\omega \sim 1/n$
with $n$ being the radial order for the mode of concern.

\subsection{Overlap with Tidal Potential}
\label{subsec:overlap}

Io orbits Jupiter in the equatorial plane with a frequency
$\omega_{\rm Io} = 2\pi/1.769\, {\rm day}^{-1}$ and at a distance $a$,
while Jupiter spins with a frequency $\Omega = 2\pi/0.413\, {\rm
day}^{-1}$. Viewed in Jupiter's rotating frame, Io rotates
retrogradely with frequency $\omega' = \Omega - \omega_{\rm Io}$ and
exerts a periodic tidal forcing on Jupiter.
We ignore Io's orbital eccentricity ($e=0.004$) in this problem. So at
a point $(r, \theta,\phi)$ inside Jupiter, the potential of the tidal
perturbation can be decomposed as
\begin{eqnarray}
\delta \Phi_{\rm Io} & = &
- {{GM_{\rm Io}}\over{a}} \left[\left({r\over a}\right)
\sin\theta \cos(\phi + \omega^\prime t) \right.
\nonumber \\
& & \left.  -  {3\over 2} \left({r\over a}\right)^2 
\left(\sin^2\theta - {1\over 3}\right) 
 - {3\over 2} \left({r\over a}\right)^2
\sin^2\theta \right.\nonumber\\
& & \left. \times \cos(2\phi + 2 \omega^\prime t)
+ {\cal O}\left({r\over a}\right)^3\right].
\label{eq:phiio}
\end{eqnarray}
The first term is necessary for maintaining the Keplerian motion of
this point in Jupiter; the second term corresponds to the potential
when Io is smeared into a ring along its orbit; the third term is the
one of relevance here. It describes the periodic forcing by Io in
Jupiter's rotating frame. Keeping only this term and writing
\be
\delta \Phi_{\rm tide}  = 
 - {{3GM_{\rm Io}}\over{2a^3}} \pomega^2 \cos(2\phi + 2 \omega^\prime
 t),
\label{eq:dphi}
\ee
we obtain $\mu_{\rm tide} = \omega_{\rm tide}/2\Omega =
2\omega'/2\Omega = 0.766$ and $m=-2$. Here, $\pomega = r \sin\theta$
is the cylindrical radius.

We investigate here the coupling between inertial-modes and the above
tidal potential. Assuming the two have the same time-dependence, we
integrate the forcing over the planet and over a period to yield the
overlap work integral,
\begin{eqnarray}
& & 
\oint dt\, 
\int d^3 r \, \rho {{\partial \boldxi}\over{\partial t}} 
\cdot \bnabla
\delta\Phi_{\rm tide}= - 
\int d^3 r\, \delta\Phi_{\rm tide} \bnabla\cdot (\rho \boldxi) 
\nonumber \\
& & = - 
\int d^3 r\, \delta\Phi_{\rm tide} \rho^\prime  =
- 
\int d^3 r\,
\delta\Phi_{\rm tide} {{\omega^2 \rho^2}\over{\Gamma_1 p}} \psi.
\label{eq:overlap1}
\end{eqnarray}
Here, $\boldxi$ and $\rho^\prime$ are the displacement and Eulerian
density perturbation from the inertial-mode, while its wave-function
$\psi$ is related to $\rho^\prime$ by $\rho^\prime = {{\omega^2
\rho}/{c_s^2}} \psi$ (eq. [9]
in Paper I). The overlap integral represents the energy pumped into
the mode per period.

\subsubsection{Tidal Overlap for the Equilibrium Tide}
\label{subsubsec:equitide}

In the limit where the tidal frequency falls well below the dynamical
frequency of the planet, the latter reacts almost instantaneously to
satisfy hydrostatic equilibrium. This tidal response is termed the
'equilibrium tide'. An extra response arises when $\omega_{\rm tide}$
has a near-resonant match with one of the free modes in the planet,
and this is called the 'dynamical tide'. Physically speaking, the
'equilibrium tide' is the sum of all the 'dynamical tide' response
driven at off-resonance.

Tidal overlap for the equilibrium tide is the largest among all tidal
response.  Disregard any time derivative in the fluid equation of
motion, take $N^2 = 0$ for the neutrally buoyant interior, and assume
any perturbation to be adiabatic,
we use equations in \S 2.1 of Paper I to obtain the following
instantaneous response,
\be
\rho^\prime_{\rm equi} = - {{\rho^2}\over{\Gamma_1 p}} \delta \Phi_{\rm tide}.
\label{eq:equirho}
\ee
The tidal overlap is,
\be
E_0 = - 
\int d^3 r\, \delta\Phi_{\rm tide}\, \rho^\prime_{\rm equi}=
{{24\pi}\over 5} \left({{GM_{\rm 
Io}}\over{a^3}}\right)^2 \int_0^R {{\rho^2 r^6}\over{\Gamma_1 p}}\, dr
\label{eq:equioverlap}
\ee
This is the energy stored in the equilibrium tide and it appears in
equation \refnew{eq:defineQ}.  Taking $M_{\rm Io} = 8.93\times 10^{25}
\g$, $a= 4.22\times 10^{10}\cm$, and adopting a Jupiter model 
from \citet{guillotreview}, we find $E_0 \approx 3\times 10^{30}
\erg$.
In comparison, the current
potential energy of Io is $\sim 3\times 10^{38} \erg$. So, over the
history of the solar system, Jupiter could have pushed Io outward for
a negligible $10^{-7}$ of its current orbit if $Q \sim 10^{13}$,
namely, only $10^{-13}$ fraction of $E_0$ is dissipated per tidal
period

As a side note, the spatial dependence of the tidal potential, as well
as that of the equilibrium tide, can be expressed in the following
form which resembles the spatial dependence of an inertial-mode in a
uniform-density sphere: $\delta \Phi_{\rm tide}
\propto \pomega^2 \propto P_2^{-2} (x_1) P_2^{-2} (x_2)$, here $x_1$
and $x_2$ are the afore-mentioned ellipsoidal coordinates. In
comparison, the lowest-order inertial-mode ($n_1 = n_2 = 0$, also
called a R-mode) has a wavefunction $\psi \propto P_3^{-2}(x_1)
P_3^{-2} (x_2)$.

\subsubsection{Tidal Overlap for Inertial-Modes}
\label{subsubsec:overlapbeta}


Consider first the tidal coupling of a gravity-mode in a solar-type
star. Firstly, this mode needs to have a spherical degree $\ell=2$ and
an azimuthal number $|m|=2$ to be compatible with the tidal potential.
Its radial eigenfunction oscillates quickly in the WKB region and
flattens out in the upper evanescent region (the convection
zone). Overlap with the (smooth) tidal potential therefore is largely
contributed by the evanescent region, with the contribution from
different nodal patches in the WKB region canceling out each other.

The situation is different for an inertial-mode. Firstly, every
even-parity, $|m|=2$ inertial-mode contains a $\ell=2$ spherical
component that can couple to the tidal potential. Moreover, the upper
evanescent region of an inertial mode is comparable in size to any
other nodal patch but with much lower density. As such it is not
particularly important for the tidal overlap.  The net tidal overlap
is the small residue after the cancellation between all regions. This
property makes it difficult to reliably calculate the overlap
integral. In fact, obtaining results in this section has been the most
difficult part of this project. Much attention is paid to ensure the
accuracy of numerical integrations, and to analytically understand the
numerical results.

We delegate much of the technical discussions to the appendixes. In
appendix \S \ref{sec:poly0}, we evaluate tidal overlap for
inertial-modes in a uniform-density model. In appendix \S
\ref{sec:polyn}, we discuss results for models of a
single power-law index ($\beta$). Lastly, in appendix \S
\ref{sec:polyn2}, we present results for models with more
realistic density profiles, including ones from Jupiter models. We
substantiate our numerical results by studying a simple toy-model
where analytical results are available. Here, we list relevant
conclusions.

We find that the severity of cancellation rises with increasing mode
order. We quantify this severity by the following dimensionless
number,
\be
{\cal C}_n =  {{
\int
\delta\Phi_{\rm tide} {{\omega^2 \rho^2}\over{\Gamma_1 p}} \psi d^3 r}\over
{ 
\int
\delta\Phi_{\rm tide} {{\omega^2 \rho^2}\over{\Gamma_1 p}} |\psi| d^3 r}}.
\label{eq:definecalc}
\ee
While ${\cal C}_n = 1$ for the equilibrium tide, ${\cal C}_n$
decreases with rising $\lambda$ (or with rising $n$ where $n= n_1 +
n_2$) with a slope that depends on the model. In detail, integration
of the top integral over the spherical angles always leads to a
cancellation of order $n^{-1}$, while integration over the radius
suffers a cancellation with a magnitude that depends on factors like
the polytropic index of the model, or discontinuities in density or
density gradient. 

As is shown in Appendix \ref{sec:poly0}, in a uniform-density sphere,
tidal overlap for all modes is zero because the material is
incompressible ($\rho' = 0$). When we adopt a constant pressure,
constant density sphere, we find that only the two lowest order
even-parity modes couple to the tide \citep{papaloizousavonije}.

For models satisfying a single power-law density profile ($\rho
\propto [1-(r/R)^2]^\beta$),  
${\cal C}_n \sim 1/n^{2\beta+1}$ for even-parity modes. For instance,
$\beta =1$ and $\beta=1.8$ yield ${\cal C}_n \sim 1/n^3$ and ${\cal
C}_n \sim 1/n^{4.6}$, respectively.  This expression is obtained from
a simple toy-model and is supported by integration of the actual
inertial-mode eigenfunctions (Appendix \ref{sec:polyn}).

Inside Jupiter, gas pressure satisfies the ideal gas law above a
radius $r/R \sim 0.98$, while it is dominated by that from strongly
interacting molecules below this radius (discussed in Appendix
\ref{subsec:density}). The density profile can be roughly fitted 
by two power-laws with $\beta$ varying from a value of $1.8$ near the
surface to $\sim 1$ deeper down. This changing $\beta$ affects the
tidal overlap. Let the transition occur over a radius $\Delta r$. We
find ${\cal C}_n \sim 1/n^3$ for $n
\leq R/\Delta r$ and ${\cal C}_n \sim 1/n^{4.6}$ for larger $n$
values. These are expected since lower order modes mostly sample the
$\beta=1$ region and are evanescent in the $\beta=1.8$ envelope, while
higher order modes experience the $\beta=1.8$ power-law.  Realistic
Jupiter models presented by
\citet{guillotreview} yield $\Delta r/R \sim 0.02$, or $\Delta r 
\sim 4$ local pressure scale heights.

The tidal overlap is also affected by discontinuities in density or
density gradient. The former may occur if, for instance, the metallic
hydrogen phase transition is of the first-order, while latter occurs
if it is of second-order.  For a density discontinuity with a
fractional value $\Delta\rho/\rho$, ${\cal C}_n \sim (\Delta
\rho/\rho) \, n^{-1}/n \propto 1/n^2$, while for a density gradient 
discontinuity of $\Delta \rho'/\rho'$, the overlap integral ${\cal
C}_n \sim (\Delta \rho'/\rho')\, 1/n^3 \propto 1/n^3$.

So in conclusion, the magnitude of the cancellation in the overlap
integral depends on the density profile, both its overall scaling with
depth as well as its interior discontinuities and sharp changes. 

In Appendix \ref{sec:polyn}, we show that one can obtain ${\cal C}_n$
by substituting the actual inertial-mode eigenfunction with a
fast-oscillating cosine function with the same number of nodes (see
Fig. \ref{fig:newpowerlaw}). It is as if one can almost make do
without detailed knowledge of the eigenfunction. This insensitivity
leads us to believe that, although we are in many cases using an
approximate solution for the inertial-mode eigenfunction, our results
for the overlap integral is reliable (more discussion in Appendix
\ref{sec:polyn2}).

Why is it necessary to go through all these detailed analysis? In the
expression for ${\cal C}_n$, while the denominator is fairly
straightforward to obtain through direct numerical integration, the
severe cancellation suffered by the integral in the numerator renders
the numerical results in many cases untrustworthy. For instance, a
$10^{-4}$ inaccuracy in the Jupiter model presents itself as a small
(but finite) density jump and affects strongly the value of ${\cal
C}_n$ at large $n$.

%

\subsection{Turbulent Dissipation}

We demonstrated in Paper I that energy of an inertial-mode is stored
mostly in the form of kinetic energy. An inertial-mode causes little
compression.
As such, its dissipation is dominated by shear viscosity.

The viscous force, ${\mbox{\boldmath $F_{\nu}$}}$, appears in the
equation of motion as
\be
\rho \ddot{\boldxi} + 2 \rho \boldOmega\times \dot{\boldxi} = 
- \bnabla p^\prime + {{\bnabla p}\over{\rho}}\, \rho^\prime - \rho
\bnabla \delta \Phi + {\mbox{\boldmath $F_{\nu}$}},
\label{eq:eqnmotion3}
\ee
where 
\be
{\mbox{\boldmath $F_{\nu}$}} = \bnabla \cdot \left(\rho \nu \bnabla
{\dot{\boldxi}}\right),
\label{eq:definef}
\ee
and $\nu$ is the shear viscosity coefficient and arises from turbulent
convection. We adopt the following mixing-length-formula
\be
\nu \sim v_{\rm cv} \ell_{\rm cv} {1\over{1+(\omega \tau_{\rm cv}/2\pi)^s}}.
\label{eq:nucv-main}
\ee
Here $v_{\rm cv}$, $\ell_{\rm cv}$ and $\tau_{\rm cv}$ are the
characteristic convection velocity, scale length and turn-over
time. When convective turn-over time is long relative to the tidal
period ($\omega \tau_{\rm cv} \gg 1$), the effective viscosity is
reduced and we adopt a reduction coefficient $s$ to describe this
behavior. We adopt $s = 2$ in our main study (see Appendix
\ref{subsec:viscosity} for more discussion) and discuss in \S
\ref{sec:discussion} the effects on our results when taking $s=1$.  We
further define the depth ($R-r$) at which $\omega\tau_{\rm cv}/2\pi =
1$ to be $z_{\rm crit}$.  For the Jupiter models we adopt (see
Appendix
\ref{subsec:viscosity} \& Fig. \ref{fig:guillot-viscosity}), $z_{\rm crit}
\approx 10^{-2.8}R$ and
\ba
\nu & \sim  & 4 (z/R)^{-1-\beta} 
\hskip0.66in {\mbox {for $z > z_{\rm crit}$}},
\nonumber \\
& \sim & 2\times 10^{10} (z/R)^{1-\beta/3}
\hskip0.2in {\mbox {for $z < z_{\rm crit}$}}.
\label{eq:nu4-main}
\ea
Here, $\beta$ is taken to be the surface value ($\beta = 1.8$). The
deeper region where $\beta=1.0$ has too weak a viscosity to be
of concern.

We assume here that the viscous forcing is small compared to the
restoring force for inertial-modes so we can ignore its effect on the
structure of inertial-modes.\footnote{This assumption is equivalent of
requiring that the rate of turbulent dissipation $\gamma$ falls much
below mode frequency $\omega$, an assumption we later confirm.} 
Viscosity does, however, dissipate mode energy. The rate of
dissipation is
\be
\gamma = {1\over E}
\int d^3 r\, {\dot{\boldxi}} \cdot {\mbox{\boldmath $F_{\nu}$}}
= - {{\int d^3 r\, \rho \nu \,\boldnabla \boldxi : \boldnabla \boldxi} 
\over {{1\over 2}\int d^3 r\, \rho\, \boldxi \cdot \boldxi}},
%
\label{eq:definegamma}
\ee
where we have integrated by part taking the surface density to be
zero. Viscosity always damps so $\gamma < 0$. In the following, we
consider only the magnitude of $\gamma$, so we re-define $\gamma =
|\gamma|$.

\subsubsection{Dissipation Rate for Equilibrium Tide}
\label{subsubsec:gammaequi}

The equilibrium tide suffers turbulent dissipation as the tidal bulge
rotates around the planet. We calculate its rate of dissipation here.

First, we obtain the displacement function ($\boldxi$) for the
equilibrium tide.  We ignore the effect of rotation here.
The motion is barotropic so $\boldxi$ is irrotational, we can write
$\boldxi = \boldnabla [f_r Y_{\ell,m}(\theta,\phi)]$, where $f_r$ is a
function of radius alone. The equation of mass conservation, combined
with equation
\refnew{eq:equirho}, yields the following equation for $f_r$:
\be
{1\over r^2} {\partial \over{\partial r}} \left(r^2 \rho {{\partial
f_r}\over{\partial r}}\right) - \rho {{\ell(\ell+1)}\over {r^2}} f_r =
b r^2 {{\rho^2}\over{\Gamma_1 p}},
\label{eq:fr}
\ee
where the tidal potential $\delta\Phi_{\rm tide} = b r^2
Y_{2,-2}(\theta,\phi)$ and $b= - \sqrt{{72 \pi}\over {15}}\, G M_{\rm
Io}/a^3$.  We solve for $f_r$ with the following boundary conditions:
near the center, the asymptotic expansion of the above equation yields
$f_r \propto r^2$, so $df_r/dr = 2f_r/r$; at the surface, the
Lagrangian pressure perturbation is zero so $\xi_r = df_r/dr = -
br^2/g$, where $g$ is the surface gravitational acceleration.

Over the whole planet, $f_r$ rises roughly as $r^2$, with a surface
tidal height $\xi_r \sim 60$ meters 
(and a comparable tangential displacement). Using the expression of
$\nu$ (eq. [\ref{eq:nu4-main}]), we obtain a damping rate of
$\gamma_{\rm equi} \approx 4\times 10^{-16} s^{-1}$.

This damping is distributed over the bulk of the planet, with roughly
equal contribution coming from each decade of depth (but little from
above $z_{\rm crit}$ where viscosity turns over). This rate can also
be estimated using $\gamma
\sim \nu/R^2$ with $\nu$ taken to be $10^4 \cm^2/\s$, the value for the
effective viscosity at the mid-point of logarithmic depth (see
Fig. \ref{fig:guillot-viscosity}). Lastly, this corresponds to an
effective Ekman number (ratio of period to viscous time-scale) of $Ek
\sim 10^{-13}$.

\subsubsection{Dissipation Rate for Inertial-Modes}
\label{subsubsec:gammainer}

Numerically, it is straight-forward to obtain the dissipation rates for
inertial-modes. It is sensitive only to the density profile at the
envelope, and is hardly affected by phase transition or other density
discontinuities in the interior. In this section, we first derive how
the rate of turbulent dissipation scales with inertial-mode wavenumber
($\lambda \approx 2(n_1 + n_2)$), and then present numerical
confirmations for these analytical scalings, using a variety of
power-law models as well as realistic Jupiter models.

We use the WKB properties of inertial-modes, discussed in
\S 3.1 of Paper I. Inertial-modes can propagate between the center and
an upper turning point, defined in the $(x_1, x_2)$ ellipsoidal
coordinates by $x_1 - \mu \sim 1/\lambda$ or $\mu - |x_2| \sim
1/\lambda$ or both. The physical depth of this turning point depends
on latitude. At $\theta \sim \cos^{-1}
\mu$ (or $x_1 \sim |x_2| \sim \mu$), it is closest to the surface with 
$z_1 \sim 2R/(1-\mu^2)/\lambda^2 \sim R/\lambda^2$ (the 'singularity
belt'); while at other latitudes, the depth is $\sim
R/\lambda$. Within the WKB cavity, the amplitude of inertial-modes
rises as $1/\sqrt{\rho}$.
In the $x_1$ and $x_2$ coordinates, nodes are spaced by $\sim
(1-\mu)/n_1$ and $\mu/n_2$, respectively, and each nodal patch (in
total $n_1\, n_2$ of these) contributes comparable amount to the total
mode energy.

Viscosity works on the gradient of the displacement. An inertial-mode
propagates with a roughly constant wavelength in most its WKB cavity,
but its wavelength shrinks drastically near or inside the singularity
belt (both $x_1 - \mu$ and $\mu - |x_2| \leq 1/\lambda$). This is
where we expect the largest dissipation to occur.  To order of
magnitude, $|\boldnabla
\boldxi| \sim \boldnabla^2 \psi \sim \lambda^2 \psi$ in the WKB cavity, 
while within the singularity belt, $|\boldnabla
\boldxi| \sim \boldnabla^2 \psi \sim  \psi/\delta_1 \sim \lambda^4 \psi$. 
We first consider modes for which $z_1 > z_{\rm crit}$,\footnote{For
Jupiter, this roughly translates to $\lambda < 50$ since $z_{\rm crit}
\sim 10^{-2.8} R$.}  so $\nu \propto z^{-1-\beta}$
(eq. [\ref{eq:nu4-main}]) in the region of interest. The work integral
of turbulent dissipation can be estimated as,
\begin{eqnarray}
& & \int d^3 r\, \rho \nu \,\boldnabla \boldxi : \boldnabla \boldxi
\nonumber \\ & \propto &
\int_{\mu}^1 dx_1 \int_{-\mu}^{\mu}
(x_1^2 - x_2^2) dx_2\, z^{-1-\beta} \rho |\boldnabla^2 \psi|^2 
\nonumber \\ 
& \propto &
\int_{\rm belt} (x_1^2 - x_2^2) dx_1 dx_2\, \rho \psi^2 \lambda^8
[(x_1^2 - \mu^2)(\mu^2 - x_2^2)]^{-1-\beta} 
\nonumber \\
& & +
\int_{\rm WKB} (x_1^2 - x_2^2) dx_1 dx_2\, \rho \psi^2 \lambda^4
[(x_1^2 - \mu^2)(\mu^2 - x_2^2)]^{-1-\beta} 
\nonumber \\
& \propto & {1\over \lambda}\, \left({1\over\lambda}\right)^2\, (\rho
\psi^2)_{z_1} \lambda^8 \lambda^{2+2\beta} \propto
\lambda^{7+2\beta}.
\label{eq:viscale}
\end{eqnarray}
Obviously, the viscous integral is dominated by the contribution from
the belt where $z \sim z_1 \sim R/\lambda^2$, and where $\theta \sim
\cos^{-1} \mu$.  Meanwhile, the mode-energy integral is dominated by
the WKB cavity with each nodal patch contributing a comparable amount,
\ba
& & \int d^3 r\, \rho \boldxi \cdot \boldxi \propto
\int_{\mu}^1 dx_1 \int_{-\mu}^{\mu}
(x_1^2 - x_2^2) dx_2\, \rho |\boldnabla \psi|^2 
\nonumber \\ 
& \sim &
\lambda^2 (\rho \psi)^2_{z_1} \left({1\over \lambda}\right)^2 n_1 n_2
\propto (\rho \psi)^2_{z_1} n_1 n_2.
\label{eq:energyintegral}
\ea
A more accurate scaling for the energy integral has been established
in Paper I (\S 3.2), yielding this integral to be $\propto
n^{2.7}\propto \lambda^{3.5}$. This latter scaling is applicable in
the range of $\lambda$ that is of interest to us and is fairly
independent of the density profile.  Returning to equation
\refnew{eq:definegamma}, we obtain
\be
\gamma \propto \lambda^{3.5 + 2\beta}.
\label{eq:analygamma}
\ee

Now we consider higher order modes for which $z_1 < z_{\rm
crit}$. Most of the damping still arises from near $z_1$, where $\nu
\propto z^{1-\beta/3}$. We repeat the scaling exercise in equation
\refnew{eq:viscale} and obtain
\be
\gamma \propto \lambda^{1.5 + 2\beta/3}.
\label{eq:analygamma2}
\ee

\begin{figure}
\centerline{\psfig{figure=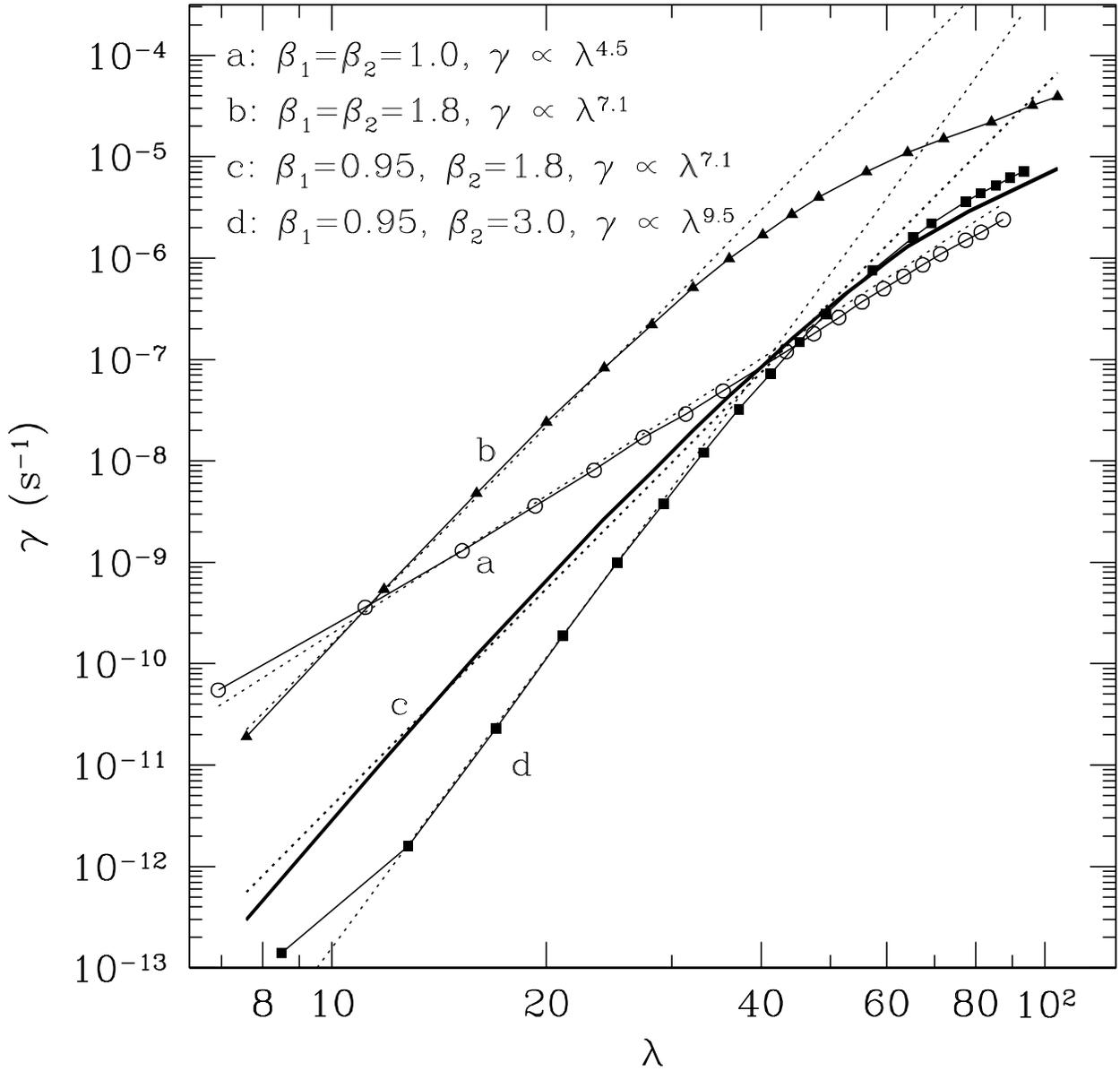,width=1.00\hsize}}
\caption{
Numerically computed turbulent damping rates for inertial-modes in
various power-law models, as a function of the mode wavenumber
$\lambda$. Here, we have taken the reduction number $s=2$ and included
damping rates only for modes with $n_1 \sim n_2$, though other modes
satisfy the same scalings observed here. Models a \& b are single
power-law models, while c \& d are double power-law models ($\beta_1$
the index in the interior and $\beta_2$ that in the envelope) with the
transition of $\beta$ occurring around
$r/R \sim 0.98$.  All models are normalized to have the same central
density and their viscosity profiles are described by equations
\refnew{eq:nu2}-\refnew{eq:nu3}. 
The dotted lines are power-law fits to the numerical results with the
numerical scalings summarized in the top-left corner. These are well
reproduced by our analytically derived relation $\gamma \propto
\lambda^{3.5+2\beta_2}$ (eq. [\ref{eq:analygamma}]). Notice that only 
the envelope power-law index enters the relation.
Damping rates in all models flatten at large $\lambda$, and scale with
$\lambda$ roughly as $\lambda^3$ (eq. [\ref{eq:analygamma2}]).
}
\label{fig:normalized-damping}
\end{figure}

In Fig. \ref{fig:normalized-damping}, we present the numerically
obtained damping rates for power-law models with $\beta$ ranging from
$1.0$ to $3.0$.  Some of these models have double power-law density
profiles but only the envelope $\beta$ value affects the scaling for
the damping rates.\footnote{For these double power-law models as well
as for realistic Jupiter models, the inertial-mode eigenfunctions are
obtained as described in Paper I.}  These numerical results confirm
our above analytical scalings.

We have also computed damping rates using realistic Jupiter models
published by \citet{guillotreview}. These models are discussed in
Appendix \ref{sec:jupiter} and have $\beta = 1.8$ in the outer
envelope. The numerical results are shown in
Fig. \ref{fig:modelBD}. They follow the scalings derived above and can
be summarized as,
\ba
\gamma & = & 6\times 10^{-13} \left({\lambda\over{7.59}}\right)^{7.1}  \hskip0.4in 
{\mbox {for $\lambda < 50$}},
\nonumber \\
& = & 3\times 10^{-9} \left({\lambda\over{7.59}}\right)^{3}  \hskip0.45in 
{\mbox {for $\lambda > 50$}},
\label{eq:actualgamma}
\ea
where we have scaled $\lambda$ by $7.59$, the value of $\lambda$ for a
low order inertial-mode ($n_1 = n_2 = 1$). Even this low order
inertial-mode is rather more strongly damped than the equilibrium
tide. Mode with $\lambda \sim 50$ have $z_1 \sim z_{\rm crit} \sim
10^{-2.8} R$. Moreover, damping rates depend only on $\lambda$ but not
on $(n_1, n_2)$ values.


\section{Tidal Q for Jupiter}
\label{sec:Qvalue}

\subsection{$Q$ value by Equilibrium Tide}
\label{subsec:Qequi}


For the equilibrium tide, equation \refnew{eq:defineQ}
\citep{goldreichsoter} yields $Q_{\rm equi} = \omega/\gamma_{\rm
equi}$,
where $\omega$ is the tidal frequency in the rotating frame ($\omega =
2\omega' = 2(\Omega - \omega_{\rm Io}) = 1.532 \Omega$), and
$\gamma_{\rm equi}$ is the turbulent damping rate for the equilibrium
tide as calculated in \S \ref{subsubsec:gammaequi}. Substituting with
the value $\gamma_{\rm equi} \sim 4\times 10^{-16} \s^{-1}$, we obtain
$Q_{\rm equi} \sim 10^{12}$, while 
\citet{goldreichphil} presented an estimate of $Q_{\rm equi} \sim
5\times 10^{13}$. The discrepancy is partially due to the fact that
they have adopted an effective $<\nu> \sim 10^3$ while our effective
$<\nu> \sim 10^4$ (\S \ref{subsubsec:gammaequi}) -- the actual
viscosity is of course uncertain, easily by a factor of $10$.
Moreover, their estimate is more order-of-magnitude in nature.
%
In any case, dissipation of the equilibrium tide, as has been argued
long and hard, can not be responsible for the outward migration of Io
and other satellites.

\subsection{$Q$ value by Inertial-Modes}
\label{subsec:Qiner}


How much stronger dissipation can inertial-modes bring about? Compared
to the equilibrium tide, inertial-modes have the advantage that they
can be resonantly driven by the tidal forcing as they are dense in the
frequency range of interest (\S
\ref{subsec:frequency}), and they are damped much more
strongly than the equilibrium tide (\S
\ref{subsubsec:gammainer}).  The disadvantage, however, lies in the
generally weak coupling between an inertial-mode and the tidal
potential. Can the first two advantages overcome the last
disadvantage? Here, we combine results from previous sections to
calculate the tidal $Q$ caused by inertial-modes.

\subsubsection{$Q$ value by Individual Modes}
\label{subsubsec:Qres}

We start by calculating the amount of tidal energy dissipated via one
inertial-mode. The following forced-damped oscillator equation
describes the interaction between an inertial eigen-mode and the tidal
forcing,
\be
\rho \ddot{\boldxi} + \rho \gamma \dot{\boldxi} 
+ \rho \omega_0^2 
\boldxi = - \rho \nabla \delta \Phi_{\rm tide} \exp^{i \omega t},
\label{eq:oscillator}
\ee
where $\boldxi$ is the displacement, and the three terms on the
left-hand-side represent, respectively, the inertia, the viscous
damping, and the restoring force. The free mode will have an
eigenfrequency of $\omega_0$. The right-hand-side is the tidal forcing
with frequency $\omega$ which we take to be $\omega \approx
\omega_0$. Adopting the substitution $\boldxi = \alpha \tilde
{\boldxi}$, $\rho^\prime = \alpha {\tilde \rho^\prime}$ with the
tilded quantities normalized as $\omega^2/2\, \int d^3 r \rho \tilde
{\boldxi}\cdot \tilde {\boldxi}= 1$, multiply both sides by ${\tilde
{\boldxi}}$ and integrating over the planet, we obtain the amplitude
$\alpha$
\be
\alpha =  {{\cal C}\over 2} {{ e^{i \omega t}}
\over {\left({{\omega_0^2}\over{\omega^2}} -1\right) +
{{2 i \gamma} \over {\omega}} }} = {{\cal C}\over 2} {{\omega e^{i
\omega t + i \epsilon}}
\over {\sqrt{4(\delta\omega)^2 + \gamma^2}}}
\label{eq:eqnalpha}
\ee
where the tidal coupling ${\cal C} = \int\, d^3 r\, {\tilde
\rho^\prime} \delta \Phi_{\rm tide}$, the frequency detuning 
$\delta\omega= \omega - \omega_0 \approx (\omega^2 - \omega_0^2)/2
\omega $, and the angle $\epsilon = \tan^{-1} (\gamma/2\delta\omega)$ 
(we assign $\gamma > 0$ for damping).  For the equilibrium tide
($\delta \omega = \omega$), this angle represents the lag-angle
between the tidal bulge and the tide-raising body, $2 \epsilon \approx
2 \tan\epsilon = \gamma_{\rm equi}/\omega = 1/Q_{\rm equi}$ 
(eq. [\ref{eq:defineQ}] \& \S \ref{subsec:Qequi}).

Energy in the inertial-mode is simply $\alpha^2$, and the energy
dissipated via this mode over one period can be found by
\begin{eqnarray}
\Delta E & = & \oint\, dt\, {{dE} \over{dt}} = \oint\, dt\, \int\, d^3 r\, 
{\rm Re}[\dot {\boldxi}] {\rm Re}[\rho \nabla \delta\Phi_{\rm tide} e^{i
\omega t}] \nonumber \\ & = & |\alpha| {\cal C} \oint\, dt\, \omega
\sin(\omega t + \delta) \cos(\omega t) \nonumber \\& = & 
|\alpha| {\cal C} \pi
\sin \delta = {{\omega {\cal C}^2 \pi \gamma} \over {2(4
\delta\omega^2 + \gamma^2)}}.
\label{eq:energylost}
\end{eqnarray}

The tidal $Q$ is related to the above quantity by eq. \refnew{eq:defineQ}
\be
Q = {{2 \pi E_0 }\over {\Delta E}} = {{4 E_0 \gamma} \over{\omega
{\cal C}^2 }} \, 
\left( {{4\delta\omega^2 +\gamma^2}\over{\gamma^2}}\right),
\label{eq:Q3}
\ee
where again $E_0$ is the energy in the equilibrium tide, and the
factor in the parenthesis describes the effect of being off-resonance.
This expression can also be derived more simply taking $\Delta E =
2\pi/\omega\, \gamma E = 2 \pi/\omega \, \gamma \alpha^2$.

We call a mode ``in resonance'' with the tide whenever
$2|\delta\omega| \leq \gamma$. The $Q$ factor associated with a
resonant mode, $Q_{\rm res}$, is proportional to the dissipation rate
and inversely proportional to the normalized tidal coupling,
\be
Q_{\rm res} = {{4 E_0 \gamma}\over{{\cal C}^2 \omega}} = {{128
\pi}\over{15}} {{\gamma}\over \omega} \, {{\left(\int {{\rho^2
r^6}\over{\Gamma_1 p}}dr\right)\,\times
\left({{\omega^2}\over 2} \int\, \rho \boldxi
\cdot \boldxi d^3 r\right)} \over {\left[ \int {{\omega^2 \rho^2
}\over{\Gamma_1 p}} \psi \pomega^2\, \cos(2 \phi)\, d^3 r\right]^2}},
\label{eq:Qres}
\ee
where equation \refnew{eq:equioverlap} is used.  Notice here that all
dependences on Io's mass and semi-major axis drop out, leaving only
the dependences on the tidal frequency and Jupiter's internal
structure.

How does $Q_{\rm res}$ behave for different inertial-modes?  Based on
our previous discussions, we introduce the following scalings with
$\lambda_0$ being the wavenumber of reference,
\ba
\gamma & = & \gamma_0 \left({{\lambda}\over{\lambda_0}}\right)^{n_\gamma}, 
\nonumber \\
{\cal C}_n & \approx & n^{- n_c} \approx \left({\lambda\over
\lambda_0}\right)^{- 1.25 n_c}.
\label{eq:scalings}
\ea
Here, ${\cal C}_n$ is the severity of cancellation in the tidal
coupling, expressed by equation \refnew{eq:definecalc}. The factor
$1.25$ in the second scaling is needed to accurately relate $n = n_1 +
n_2$ to $\lambda$ in the range of interest. The normalized tidal
coupling ${\cal C}$ can be related to ${\cal C}_n$ as
\ba
{\cal C} & = & {\cal C}_n\, {{\int
\delta\Phi_{\rm tide} {{\omega^2 \rho^2}\over{\Gamma_1 p}} |\psi| d^3 r}\over
{\left({{\omega^2}\over 2} \int \rho \boldxi
\cdot \boldxi d^3 r\right)^{1/2}}} \nonumber \\
& & \propto {\cal C}_n \, {{(\sqrt{\rho} |\psi|)_{z_1}} \over
{{(\sqrt{\rho} |\psi|)_{z_1}} (n_1 n_2)^{1/2}}} \nonumber \\
& & \propto 
{{{\cal C}_n}\over n} 
 \propto {{{\cal C}_n}\over \lambda}.
\label{eq:relateCCn}
\ea
Here, we have used the information that the envelope of $\psi$ scales
as $1/\sqrt{\rho}$ in the WKB region, and that every nodal patch in
the WKB region contributes comparable amount of kinetic energy to the
total budget. Again $z_1$ stands for the upper turning point at 
latitude $\theta = \cos^{-1} \mu$.

These scalings combine to yield the following expression for $Q_{\rm
res}$:
\be
Q_{\rm res} = Q_0 \gamma_0 \left({{\lambda}\over{\lambda_0}}\right)^{n_Q}
= Q_0 \gamma_0
\left({{\lambda}\over{\lambda_0}}\right)^{n_\gamma + 2 + 2.5 n_c},
\label{eq:Qresscaling}
\ee
where $Q_0$ is a constant that depends on Jupiter's internal structure.

\begin{figure}
\centerline{\psfig{figure=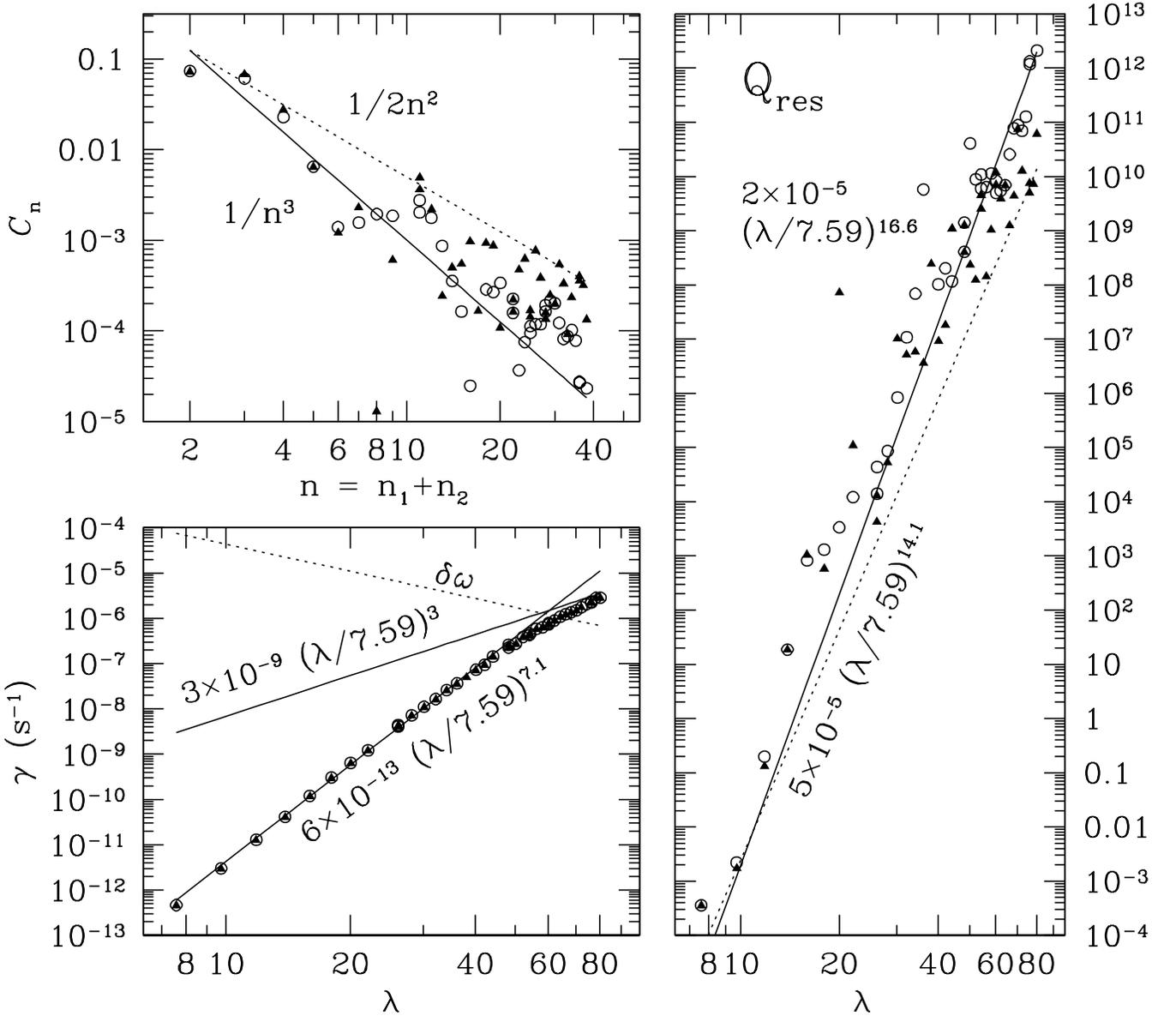,width=1.00\hsize}}
\caption{
Tidal coupling, viscous damping rate and resonant $Q_{\rm res}$ factor for
various inertial-modes calculated using two Jupiter models.
Model B (open circles) has no heavy metal core and no first-order
metallic hydrogen phase transition, while model D (solid triangles)
has a core as well as a density jump at $r/R \sim 0.8$ due to the
plasma phase transition.  The upper-left panel presents the
(normalized) coupling integral ${\cal C}_n$
(eq. [\ref{eq:definecalc}]) as a function of inertial-mode node
numbers ($n = n_1 + n_2$). Although the scatter is large, model B
results are best fit by ${\cal C}_n \sim 1/n^3$ (solid line), while
model D results follow ${\cal C}_n \sim 1/2n^2$ (dotted line). The
lower-left panel shows the energy damping rate as a function of mode
wavenumber $\lambda$ ($\lambda \sim 2 n$). Results from both models
scale as $\lambda^{7.1}$ for low-order modes and as $\lambda^3$ for
high-order modes (two solid lines), consistent with analytical
expectations (\S \ref{subsubsec:gammainer}).  The dotted line in the
same panel is the minimum frequency detuning as a function of
$\lambda$ (eq. [\ref{eq:bestdmu}]).
$Q_{\rm res}$, the $Q$ value contributed by each mode when it is in
resonance with the tide (eq. [\ref{eq:Qres}]), is plotted on the
right-hand panel as a function of $\lambda$. Again, analytical
expectations for models B \& D are depicted by the solid and dotted
lines, respectively. While low-order modes ($\lambda < 40$) from the
two models largely share similar $Q_{\rm res}$ values, higher order
modes follow more closely the analytical scalings.  Here, we have
included only inertial-modes with $\mu \sim 0.776$ but the results
remain similar for other inertial-modes.}
\label{fig:modelBD} 
\end{figure}

We obtain numerical results using two realistic Jupiter models
published by \citet{guillotreview}: models B and D. They are discussed
in detail in Appendix \ref{sec:jupiter}. Of particular relevance is
that, while hydrogen metallic phase transition is treated as a smooth
transition in model B (interpolated equation of state), model D has a
first-order phase transition and the associated density jump occurring
around $r/R \approx 0.8$. As a result, ${\cal C}_n \sim 1/n^3$ ($n_c =
3$) in model B,\footnote{Two factors contribute comparably to this
scaling: the sharp transition of equation of state near $r/R \sim
0.98$ and the discontinuous density gradient at the phase transition
point.}  while ${\cal C}_n \sim 1/2n^2$ ($n_c = 2$) in model D. These
scalings are derived analytically in Appendix \ref{sec:polyn2}, and
tested using a toy-model integration. In Fig. \ref{fig:modelBD}, we
further demonstrate that these scalings indeed apply to
inertial-modes, albeit with quite a bit of fluctuations.

From equation \refnew{eq:actualgamma}, we obtain $n_\gamma = 7.1$ for
low-order modes ($\lambda < 50$). So $n_Q$ is expected to be $16.6$
for model B and $14.1$ for model D. We present numerically calculated
$Q_{\rm res}$ in Fig. \ref{fig:modelBD} and they confirm these
scalings.  Moreover, $Q_{\rm res}$ ranges from $10^{-4}$ for the
lowest order inertial-modes to $10^{10}$ for model B (and $10^8$ for
model D) when $\lambda \sim 50$.

\subsubsection{Overall $Q$ Value}
\label{subsubsec:overallQ}

If we consider multiple inertial-modes each causing $Q_i$, the total
effect is
\be
Q = {1\over {\sum_i 1/Q_i}}.
\label{eq:Q4}
\ee
So at any given tidal frequency, $Q$ is dominated by the mode that
contributes the smallest $Q_i$. Which mode is this and what is the
resulting $Q$ value? We derive analytical scalings here to answer
these questions.

At a given forcing frequency, $Q$ values (eq. [\ref{eq:Q3}]) for
different modes depend on $\lambda$ non-monotonically. Typically, as
$\lambda$ increases, $Q$ first decreases and then rises sharply. This
is because low-order modes typically are driven off-resonance
($2|\delta\omega| \geq \gamma$) while one can easily find high order
modes to be in resonance with the tide. For low-order modes, as
$\lambda$ increases, the chance for a good resonance with the tidal
frequency improves. This compensates for the fact that tidal coupling
weakens with $\lambda$ and
\be
Q \approx Q_{\rm res} {{4\delta\omega^2}\over{\gamma^2}} \approx {{256
\pi^2 \Omega^2 Q_0}\over{\gamma_0 \lambda_0^4}} \left({\lambda \over
\lambda_0}\right)^{n_Q - 2 n_\gamma -4}.
\label{eq:anotherlimit}
\ee
Here $n_Q - 2 n_\gamma - 4 = 2.5 n_c - 2 - n_\gamma < 0$. For
high-order modes that satisfy $2|\delta
\omega| \leq \gamma$, increasingly
weaker tidal coupling accounts for the fact that $Q$ rises with
$\lambda$ as $Q \approx Q_{\rm res} \propto \lambda^{n_Q}$.
The lowest $Q$ value is to be found around modes that satisfy
$2|\delta \omega| = 2|\omega_0 - \omega| \sim \gamma$. This occurs at
\be
{\lambda \over \lambda_0} \approx \left({{16 \pi \Omega}\over{\gamma_0
\lambda_0^2}}\right)^{1/(n_\gamma+2)}.
\label{eq:lambdaeq}
\ee
For Jupiter models, this yields $\lambda \sim 60$ (also see
Fig. \ref{fig:modelBD}). These are the modes that are most relevant
for tidal dissipation. They give rise to a minimum $Q$ value 
\be
Q \approx Q_0 \gamma_0 \left({{16 \pi \Omega}\over{\gamma_0
\lambda_0^2}}\right)^{n_Q/(n_\gamma+2)}. 
\label{eq:miniQ}
\ee
This roughly corresponds to $Q \sim 10^{10}$ for model B, and $Q \sim
10^8$ for model D (see more detailed calculation below).



In the following, we confirm and refine the above analytical results
by a numerical model. While we have a reasonably good handle on mode
damping and tidal coupling, we do not have a perfect Jupiter model nor
exact inertial-mode solutions to produce exact mode frequencies.  So
we could not reproduce exactly the tidal response of Jupiter as a
function of Io's orbital period.  Fortunately, this problem can be
circumvented.  In the following exercise, we produce an artificial
spectrum of inertial-modes, with frequencies that satisfy the WKB
dispersion relation $\mu = sin(n_2
\pi/\lambda)$ with $\lambda
\approx 2(n_1 + n_2)$ (Paper I). To this frequency we add a small
random component of order $\delta\mu=\mu/(10 n_1)$, which is of order
$1/10$ the frequency spacing between neighboring modes of the same
$n_2$ value. This random component is to encapsulate our above
ignorance
but neither its size nor its sign qualitatively affect our conclusion.
In this treatment, although we will not be able to obtain the exact
tidal response of the planet at each forcing frequency, we can get a
reasonable statistical impression. In fact, this is the only logical
approach warranted by our current knowledge of the interior of
Jupiter.

For each inertial-mode, we assign a damping as in equation
\refnew{eq:actualgamma}, and a $Q_{\rm res}$ as in equation
\refnew{eq:Qresscaling} (different for models B \& D), 
and calculate, as a function of tidal frequency, the $Q$ value for
individual modes as well as the overall $Q$ value. The results are
presented in Figs. \ref{fig:Qjupiter-B} \&
\ref{fig:Qjupiter-D} for the two models. One observes that the
$Q$ value fluctuates wildly as a function of the forcing
frequency. While there is a ceiling to the overall $Q$ value, there
may be occasions when resonance with very low-order modes occurs,
leading to deep valleys with $Q$ reaching values as small as $10$. The
ceiling, on the other hand, is determined by $\lambda \sim 60$ modes
which are always in resonance at any forcing frequency. Due to their
high $Q_{\rm res}$ values, modes of orders higher than these are not
important.

The results should be interpreted statistically.  One can infer from
them two pieces of information about Jupiter's $Q$ value. The first is
the average $Q$ value across a certain frequency range, and the second
the probability of $Q$ value falling below $10^6$ in this frequency
range. Here, the value $Q = 10^6$ is taken to be the rough upper limit
for the empirically inferred $Q$ value.

The definition for the word 'average' deserves some deliberation.  We
follow \citet{goodmanoh} and \citet{terquem} in adopting the following
average,
\be
{\bar Q} \equiv {{\int_{\mu_1}^{\mu_2} Q(\mu) Q(\mu) d\mu}\over{
\int_{\mu_1}^{\mu_2} Q(\mu) d\mu}}.
\label{eq:barQdef}
\ee
This is equivalent to a time-weighted average since the time a system
spends in a certain state is inversely proportional to the torque at
that state. Over the evolutionary timescale, the system quickly moves
through the deep valleys (large torque) and lingers around in the
large $Q$ region. This is also where we most expect to find Jupiter
today.

We find that for $\mu \in [0.7,0.8]$, ${\bar Q} \approx 1.4 \times
10^9$ for model B and ${\bar Q} \approx 5.8\times 10^7$ for model D,
roughly consistent with our analytical estimates. Recall that $Q_{\rm
equi} \approx 10^{12}$. Moreover, at any forcing frequency, 
the probability that Jupiter has $Q < 10^6$ is $3\%$ in model B and
$\sim 10\%$ in model D.


\section{Discussion}
\label{sec:discussion}

Throughout our calculation, we have assumed that Jupiter is uniformly
rotating, neutrally buoyant and core-less. We have also assumed that
its internal convection provides a turbulent viscosity which is
quantified by the mixing length theory and which is reduced with an
index $s=2$ when the convection turn-over time is long compared to the
tidal period. We obtained inertial-mode eigenfunctions for realistic
Jupiter models using a combination of WKB approximation and exact
surface solution (Paper I).

In this section, we discuss the validity of our various assumptions,
factors that might influence our results, as well as implications of
our results.

\subsection{Tidal Overlap}
\label{subsec:uncertain}

Firstly, a precaution about tidal overlap. We find that this is the
trickiest part of our work because inertial-modes propagate
essentially over the whole planet, with a small evanescent region very
close to the surface. Regions of positive and negative tidal coupling
lay side by side, leading to strong cancellation and extreme
sensitivity to numerical accuracy. In fact, for a sphere with a
density profile that follows a single power-law, the net tidal
coupling decreases with increasing mode order so strongly (Appendix
\ref{sec:polyn}) that numerical precision is soon strained even for fairly 
low-order modes. Inertial-modes are not important for tidal
dissipation in these models.

In a realistic Jupiter model, the cancellation is less extreme due to
the following two features: the molecular to metallic hydrogen
transition at $r/R \sim 0.8$ (either a discreet phase transition or a
continuous change) and the polytropic index change at $r/R \sim 0.98$
where hydrogen molecules change from ideal gas to strongly interacting
Coulomb gas (discussed in Appendix
\ref{subsec:density}). These two features act as some sort
of 'internal reflection' for the inertial-modes -- their WKB envelopes
inside and outside of these features differ. This weakens the
above-mentioned near-perfect cancellation in the overlap contribution
from different regions and leads to larger tidal coupling. This is
confirmed by integration using both a toy-model (Appendix
\ref{sec:polyn2}) and actual inertial-mode eigenfunctions. In this
case, tidal dissipation via inertial-modes outweighs that due to the
equilibrium tide.

The inertial-mode eigenfunctions for realistic Jupiter models are
constructed as follows (see also Paper I). We first obtain
eigenfunctions for a single power-law model with the power-law index
($\beta$) determined by that in the outer envelope of the Jupiter
model. This can be done exactly as long as we ignore the Eulerian
density perturbation in the equation of motion.\footnote{This term is
small and its removal from the equation of motion, as we discussed in
Paper I, does not preclude tidal forcing between the tide and the
inertial-modes.}
We multiply the resulting wave-function by a factor $\sqrt{\rho_{\rm
surf}/\rho}$ where $\rho_{\rm surf}$ is the density for the above
single power-law and $\rho$ is the actual density. We showed in Paper
I that in the WKB region, this construction approximates the actual
eigenfunction to order ${\cal O}(1/\lambda^2)$, and it is exact in the
surface evanescent region.

The effects of such a non-exact formulation on mode eigenfrequencies
will not significantly alter our results and its effects on the
damping rates are negligible.  But does it affect our results on tidal
coupling, which, as we have shown, depends sensitively even on
numerical accuracy? A definitive answer may have to come from high
resolution numerical calculations. But our toy-model gives us some
confidence that our approach has captured the essence of the problem
and that our overlap result is qualitatively correct.



\subsection{Turbulent Viscosity}
\label{subsec:uncertainvis}

The next issue concerns the turbulent viscosity. We have presented the
detailed viscosity profile in Appendix \ref{subsec:viscosity} \&
Fig. \ref{fig:guillot-viscosity}.. This is calculated based on the
mixing-length theory which is order-of-magnitude in nature (also see
eq. [\ref{eq:nucv-main}]). How much does the $Q$ value change when the
viscosity is raised (or decreased) by a factor of, say, $100$? The
scaling in equation \refnew{eq:miniQ} yields $Q \propto
\gamma_0^{1-n_Q/(n_\gamma+2)}$, ot $Q \propto \gamma_0^{-0.8}$ for model B and 
$\propto \gamma_0^{-0.5}$ for model D. So even a factor of $100$
change in the viscosity causes a change in the $Q$ value that is
comparable to our numerical accuracy and is not significant.

\citet{zahn} has advocated a less drastic reduction of the
turbulent viscosity when the convective turn-over time is much longer
than the tidal period: $s=1$ in equation
\refnew{eq:nucv-main}. This produces two differences to our results. First, the
equilibrium tidal $Q$ is reduced to $\sim 10^9$ as the effective
viscosity is increased over the bulk of the planet by a factor of
$\sim 10^3$.
Inertial-modes also in general experience stronger dissipation, with
the change more striking for low-order modes. Moreover, modes of lower
order can now satisfy the resonance condition ($2|\delta
\omega| \approx \gamma$) and they are the dominant modes for tidal dissipation.
However, the enhanced $\gamma$ also means every mode now has a larger
$Q_{\rm res}$, as a result, the overall $Q$ factor by inertial-modes
is hardly modified from that in the $s=2$ case (see
Fig. \ref{fig:Qjupiter-Ds1}).

\subsection{Density Discontinuities}
\label{subsec:firstorder}

\begin{figure}
\centerline{\psfig{figure=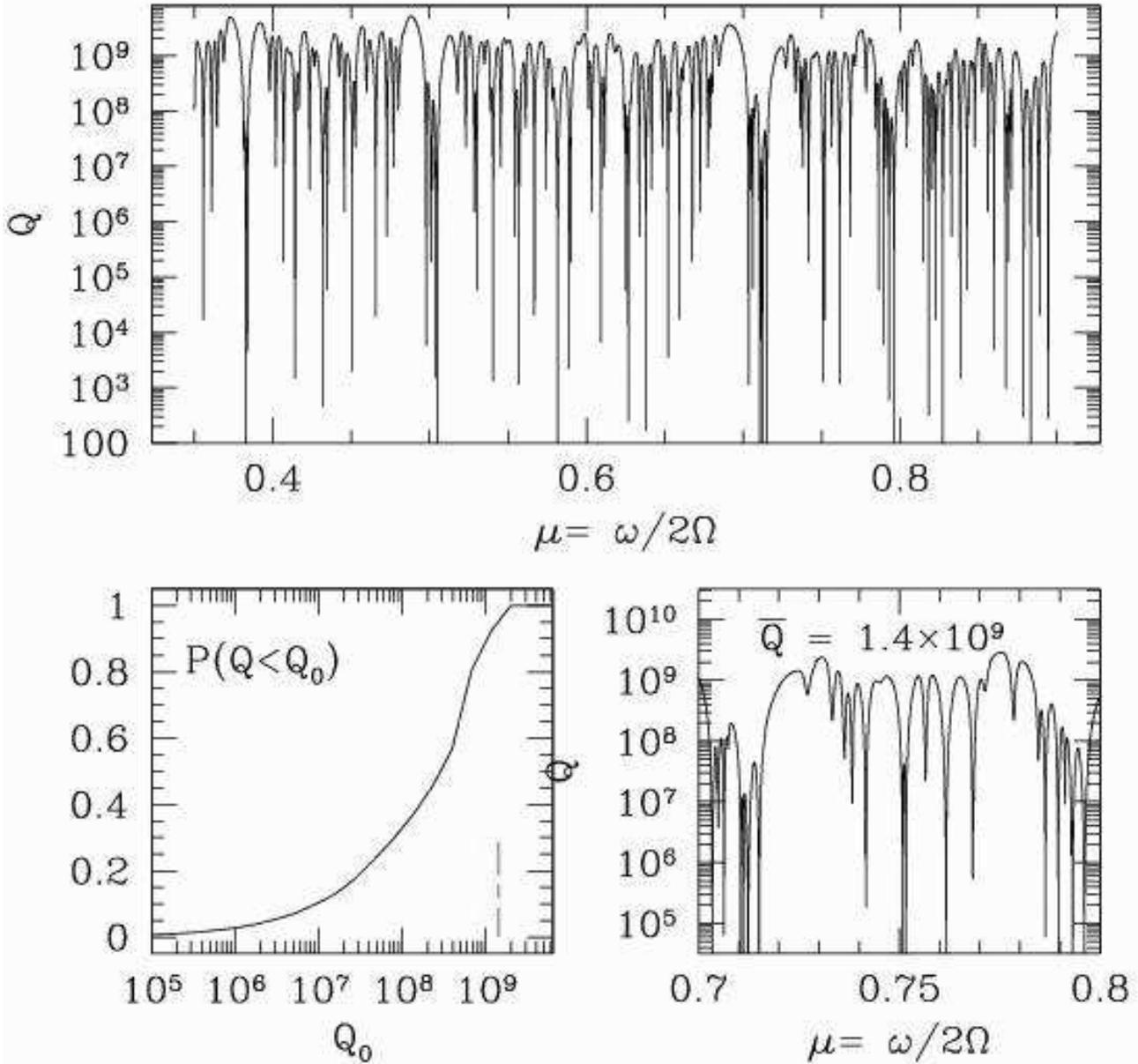,width=1.00\hsize}}
\caption{
Numerically calculated values of $Q$ for model B.  The upper panel
shows $Q$ as a function of the tidal frequency in the rotating frame.
Deep dips occur whenever the tide is in resonance with a low-order
inertial-mode ($\lambda \ll 60$), and the ceiling to the $Q$ value is
determined by the group of modes with $\lambda \sim 60$ which satisfy
$2|\delta\omega| \sim \gamma$. The tide is always in resonance with
one of these modes at any frequency.
The lower left panel shows the cumulative probability distribution of
the $Q$ value within the frequency range $0.7 < \mu < 0.8$. At a given
frequency 
there is $\sim 3$ percent chance that we will find $Q < 10^6$. The
probability for this to occur at a given instant in time is smaller.
The dashed vertical curve locates the time-weighted average $Q$ value
(${\bar Q}$, eq. [\ref{eq:barQdef}]). We find ${\bar Q} = 1.4\times
10^9$ within this frequency range.  The lower right panel expands the
view of the upper panel over this frequency range.  The locations of
the fine structure in this plot are not to be taken literally as we
have adopted an approximate dispersion relation for the
inertial-modes.
 }
\label{fig:Qjupiter-B}
\end{figure}

As our results in Figs. \ref{fig:Qjupiter-B} \&
\ref{fig:Qjupiter-D} show, when there exists a discreet density jump
inside Jupiter, the overall $Q$ factor is $\sim 10^7$, or $\sim 10^2$
times smaller than the case when there is no jump, with $\sim 10\%$
chance that the current $Q$ value falls between $10^5$ and $10^6$ (the
empirically inferred $Q$ range for Jupiter). This dependence on
density discontinuity deserves explanation.

It results from a difference in the overlap integral. In the jump
case, cancellation in the overlap contribution coming from different
parts of the planet is less severe (${\cal C}_n
\propto 1/n^2$), while it is more complete in the no-jump case (${\cal
C}_n \propto 1/n^3$), as is explained using a toy-model in Appendix
\ref{sec:polyn2}.  
In the no-jump case, the ${\cal C}_n \propto 1/n^3$ scaling may arise
from two causes: a discontinuity in the density gradient due to, for
instance, a second-order phase transition, and a sharp transition in
the power law index $\beta$ (equivalently, the polytropic index
$\Gamma_1$) when the equation of state changes. In Jupiter models, the
latter occurs at $r/R \sim 0.98$, spanning a range of $\Delta r/R \sim
0.02$, or $\sim 4$ local pressure scale heights (Appendix
\ref{subsec:density}). The overall $Q$ factor is little
affected if either transition region is shifted upward or downward by
a few pressure scale heights. However, if the second-order phase
transition does not exist, {\it and} if the polytropic transition
occurring over a range $\Delta r/R \gg 0.02$, we expect ${\cal C}_n
\propto 1/n^{4.6}$ and the overall $Q$ factor to be much larger.

Does Jupiter harbor a density jump?

One possibility is the so-called metallic hydrogen phase
transition. Our knowledge of the equation of state for hydrogen at
Mbar level is currently limited. We do not know whether the transition
from a molecular fluid to a conductive fluid (metallic hydrogen) is a
plasma phase transition (PPT) with a discreet density jump, or a
continuous process with only a jump in the density gradient. And in
the case of PPT, we do not know whether the actual Jovian adiabat
falls below or above the critical temperature for a first-order
transition (Stevenson, private communication). Plighted by these
uncertainties, planet modelers have typically chosen to insert (or not
to insert) by hand a small density jump at the suspected PPT location,
and then interpolated between very low and very high pressures (where
we know the equation of state well), under certain assumptions, to
obtain the pressure-density curves around this point. We build our
analysis on two examples of such models (model B with a smooth
transition and model D with a jump). Interestingly,
\citet{guillotreview} showed that among models that match all
observational constraints on Jupiter, the ones with PPT equation of
state have larger core mass and lower total mass of heavy elements,
while the ones with smooth interpolated equation of state tend to the
opposite.

Another possibility may follow from helium/hydrogen phase separation.
Whenever the Jovian adiabat falls below the critical temperature curve
for helium immiscibility, helium separates from hydrogen and forms
helium-rich droplets that fall toward the center \citep{salpeter}. Due
to its cooler interior, this process has proceeded further in Saturn
than in Jupiter. But even in Jupiter there may be a density jump, or
at worst, a jump in density gradient, associated with this effect.

Close-in hot exo-jupiters presumably have higher overall entropy than
Jupiter does, as radiation from their surface is effectively sealed
off by the stellar insulation. Their interior temperature is higher at
a given pressure. Both PPT and helium rain-out are therefore less
likely to occur in these bodies.

In summary, current Jupiter models exhibit features that warrant $Q
\sim 10^9$. It is plausible to find a non-negligible density jump 
in the Jovian interior, due either to a first-order PPT or
helium/hydrogen separation, in which case we obtain $Q\sim
10^7$. This, however, is more difficult to justify in hot
exo-jupiters, compromising our initial goal of searching for a
universal mechanism.

\subsection{Presence of a Solid Core}
\label{subsec:core}

We have assumed here that convection penetrates into the center of
Jupiter. But it is possible that Jupiter does have a solid core.
\citet{dermott} pointed out that body tide in the
(imperfectly elastic) solid core of Jupiter with a core quality factor
$\sim 30$ can account for the observed tidal dissipation.
However, this requires a core size which is at the upper-end of
current determinations ($r_{\rm core}/R \sim 0.15$) as well as a core
quality factor which is currently unknown. 
Moreover, the efficiency of such a mechanism depends sensitively on
the core size and it may be unreasonable to expect that exo-jupiters
all have core sizes within a narrow range.
So we restrict ourselves to consider the effect of a core on the tidal
$Q$ factor due to inertial-modes. 

Inertial-modes are excluded from the solid-core.  For an estimate, we
retain the inertial-mode eigenfunctions calculated for the core-less
case, but suppress from the core region contribution to mode energy,
mode damping, and tidal overlap integral. We find no substantial
difference between this and the core-less case (one can also compare
results from model B which is core-less and model D which has a $10
M_\oplus$ core). Contribution from the core region to the overlap
integral, for instance, is insignificant as the radial integrand drops
as $\propto r^6$ (eq. [\ref{eq:reduction}]): radial dependence of the
tidal potential goes as $r^2$, and inertial-modes are more anelastic
(small $\rho^\prime$) in the high density region. 

A more subtle influence of the core, however, may be present. While we
have been able to separate spatial variables and calculate
inertial-mode eigenfunctions in the ellipsoidal coordinates for
core-less models, the presence of a spherical core destroys this
convenience. The inner boundary conditions
can no longer be defined along constant ellipsoidal coordinate curves
and we have to return to the original partial differential
equations. This is analogous to the situation where the Coriolis force
breaks the symmetry of a spherical star, with the result that the
angular dependence of an eigen-mode in a rotating star can no longer
be described by a single spherical harmonic but only by a mixture of
them.  So it is perceivable that, if we adopt core-less inertial-mode
eigenfunctions as a complete basis, inertial-mode eigenfunction in the
presence of a spherical core may be a mixture of these functions. This
gives us a hint on how to proceed when there is a core. It is possible
to obtain the mixing ratio and use these to calculate new damping
rates, mode energy and tidal coupling. We conjecture that the mixture
becomes purer (more dominated by one component) as the core size
approaches zero. In particular, we expect the mixing not to be
important when the core size is much smaller than a wavelength of the
inertial-mode ($r_{\rm core}/R \ll 1/\lambda$). We plan to extend our
calculation to the solid core case in the future.

The above conjecture seems to be supported by numerical calculations
by \citet{gordon2}. He recovers low-order inertial-modes when he
decreases the core size. When the core size is significant, however,
OL's study discovered something else. Instead of global
inertial-modes, they found that fluid response to the tidal forcing is
concentrated into characteristic rays which become singularly narrow
as viscosity goes to zero.  This appears a rather different picture
from ours and the physical origin of these singular rays deserves
understanding.

\subsection{Radiative Atmosphere}
\label{subsec:radiative}

We have also assumed that the convection zone extends all the way to
zero density. This may be unrealistic for Jupiter, and worse still for
exo-jupiters. In the Jupiter models we adopted, convection gives way
to radiation just above the photosphere ($p = 1 bar$). 
The reality is more complicated (also see discussions in Paper
I). Temperature in the Jovian atmosphere is such that as a fluid
parcel travels upward, its water content condenses and releases latent
heat. The resulting adiabatic gradient (the 'wet adiabat') depends on
the water content and is shallower than the one that does not include
water condensation (the 'dry adiabat').
So for a given temperature profile, a particularly dry parcel can be
convectively stable. This is consistent with the Galileo probe data
which indicates stable stratification down to $\sim 20 bar$ after
entering a dry spot on Jupiter\citep{allison}. Available Jupiter
models are at best 1-D representation of the 3-D structure, and our
results depend critically on the temperature structure and turbulent
viscosity in the upper atmosphere of Jupiter. 

What is the effect of a thin radiative atmosphere on inertial-modes? 
Inertial-modes may not be perfectly reflected near the surface and
some of its wave-flux can be smuggled out of the convective region in
the form of gravity-waves.  The radiative zone has a peak \Bruntfreq
buoyancy frequency
\be
N \sim {g \over {c_s}}\sim {{2700} \over{ 9.3\times 10^4}} \sim 0.029 \, \s^{-1},
\label{eq:atmN}
\ee
which is much higher than the inertial-mode frequencies we are
interested in ($\omega \sim 3.5\times 10^{-4} \s^{-1}$). So the
relevant gravity-wave is high in radial order and is strongly modified
by rotation, satisfying $N \gg \omega \sim \Omega$. Such waves can be
calculated (semi)-analytically under the 'traditional approximation'
and are called the 'Hough modes'.  The smuggled wave-flux is
subsequently lost in the higher atmosphere where the gravity-wave
breaks. This brings about enhanced damping to the
inertial-mode. Recall that the overall $Q$ factor scales roughly as
inverse square root of the damping rate. So unless the resultant
damping rate is orders of magnitude above the rate of turbulent
damping, the overall $Q$ factor is little affected.

There are other ways in which a radiative envelope may affect
inertial-modes. The upper-turning point ($z/R \sim 1/\lambda^2$ when
$\theta \sim \cos^{-1} \mu$ and $z/R \sim 1/\lambda$ otherwise) of a
sufficiently high order inertial-mode may fall near or above the
convective-radiative interface. When this occurs, the structure of the
inertial-mode is significantly modified. The radiative region imposes
a different surface boundary condition on the inertial-mode than the
one we assume here (vanishing Lagrangian pressure perturbation). This
different boundary condition, as is illustrated by the toy model in
Appendix
\ref{sec:polyn2}, may give rise to much different (likely larger) tidal overlap and
therefore a different $Q$ (likely smaller) factor (see also \S
\ref{subsec:previouswork}). 


Extra-solar hot jupiters are strongly irradiated by their host
stars. Their atmosphere is more isothermal leading to a substantially
thicker radiative envelope (down to $\sim 30\km$ below photosphere)
than that in Jupiter. This envelope may sustain rotationally-modified
gravity-waves ('Hough Modes') which may be resonantly (if these waves
are trapped) excited by the tidal potential. It is possible that this
explains the tidal dissipation in these hot jupiters
\citep{lubow}. However, inertial-modes should still exist and will couple
to the tidal potential even in these planets.  The fact that the
$Q$-values appear to be similar between the exo-jupiters and our
Jupiter leads us to suspect that inertial-modes will remain
relevant. It is foreseeable, for instance, that these planets harbor a
new branch of global modes which are inertial-mode like in the
interior and gravity-mode like in the exterior.

\subsection{Where is the tidal energy dissipated?}
\label{subsec:whichlayer}

In our picture of resonant inertial-mode tide, most of the tidal
dissipation occurs very near the surface, where both the kinematic
viscosity and the velocity shear are the largest. In a realistic
Jupiter model, the effective turbulent viscosity peaks at a depth of
$\sim 60 \km$ ($z_{\rm crit}$, Fig. \ref{fig:guillot-viscosity} in
Appendix \ref{subsec:viscosity}), and decays sharply
inward. Meanwhile, the displacement caused by inertial-modes rises
outward toward the outer turning point.
And the velocity shear reaches its maximum inside the 'singularity
belt' (Paper I), which is found to be around $\theta
\approx \cos^{-1} \mu$, with  an angular 
extent $\sim R/\lambda$ and a depth $\sim R/\lambda^2$. For
inertial-modes most relevant for tidal dissipation ($\lambda \sim
60$), this depth roughly coincides with the location of maximum
viscosity. We have confirmed numerically that most of the dissipation
indeed occur in this shallow belt.

The tidal luminosity in Jupiter is
$\sim 7\times 10^{20} (10^6/Q) \erg/\s$. What is the effect of
depositing this much energy in a shallow layer? We compare this
against intrinsic Jovian flux of $F \sim 5000 \erg/\cm^2/\s$. The
total intrinsic luminosity passing through the belt is $\sim 2\pi
R^2/\lambda F \sim 3\times 10^{22} \erg/\s$.  This is larger than (or
at worst comparable to) the tidal luminosity. Another way of phrasing
this is to say that the local thermal timescale is shorter than (or at
worst comparable to) the ratio between local thermal energy and the
tidal flux. So the belt is expected to be able to get rid of the tidal
energy without suffering significant modification to its structure.

Angular momentum is also deposited locally. We assume here that the
convection zone is able to diffuse the excess angular momentum almost
instantaneously toward the rest of the planet. However, if convective
transport is highly anisotropic and prohibits diffusion, it is
possible that this (negative) angular momentum is shored up near the
surface and contributes to surface meteorology of Jupiter. 


The transiting planet HD209458b is observed to have a radius of $\sim
1.3 R_J$ \citep{brown}. Its proximity to its host star and its
currently near-circular orbit raise the possibility that its over-size
is a result of (past or current) tidal dissipation
\citep{gu}. However, if our theory applies also to these hot jupiters,
we would expect that the tidal heat is deposited so close to the
planet surface that it can not be responsible for inflating the
planet.\footnote{It is difficult to imagine how entropy deposited near
the surface can be advected inward to raise the entropy level of the
entire planet.} 
Moreover, given the short local thermal timescale, any change to the
planet structure should disappear once tidal dissipation ceases.
%

\subsection{Tidal Amplitude and Nonlinearity}
\label{subsec:amplitude}

If inertial-modes are resonantly excited to large amplitudes, they can
transfer energy to other inertial-modes in the planet and be
dissipated by nonlinear mode coupling. To see whether this is
important, we consider the amplitude of inertial-modes. This is
largest near the surface around the 'singularity belt'. When an
inertial-mode is resonantly excited ($|\delta\omega| \leq \gamma$), we
obtain a horizontal surface displacement $\xi_h \sim 10^{11}
(\lambda/7.59)^{-7} \cm$. While this implies extreme amplitudes for
low-order modes, they only come into resonance rarely. For modes of
interest ($\lambda \sim 60$), the typical surface displacement
amplitude is $\sim 10^3 \cm$,\footnote{In contrast, the displacement
amplitude of the equilibrium tide is much larger, $\xi_h \sim 60
\m$. $Q_{\rm equi}$ is large, however, because the equilibrium tide is 
dissipated very weakly.} so the dimensionless amplitude ($\xi/R_J$) is
$10^{-7}$. Can such an amplitude incur strong nonlinear damping?

At such small amplitudes, nonlinear effects can be well described by
three-mode couplings. The efficiency of this process scales with the 
amplitudes of the modes concerned.
The most important nonlinear coupling is parametric resonance: when the
inertial-mode reaches a threshold amplitude, pairs of
daughter inertial-modes, at half the frequency and with $m=-1$, 
can be parametrically
excited and can grow to significant amplitudes. Nonlinear mode 
coupling then drains energy quickly out of the original mode. The
threshold dimensionless amplitude is \citep{landau,wugoldreich}
\be
\left.{\xi \over {R_J}}\right|_{\rm para} \approx {1\over \kappa}
\left[\left({{\gamma_2}\over{\omega}}\right)^2 
+ \left({{\delta\omega}\over{\omega}}\right)^2\right]^{1/2},
\label{eq:parametric}
\ee
where $\kappa$ is the coupling coefficient between the parent and the
daughter pair, $\gamma_2$ the damping rate for the daughter modes, and
$\delta\omega$ the frequency detuning for this resonance.
\citet{arras} has studied the coupling coefficient for inertial-modes in a
uniform density sphere and found $\kappa \leq n_1/\mu^2$: the maximum
coupling coefficient obtains for daughter pairs that are spatially
similar and maximally overlap.\footnote{In their normalization, 
the dimensionless amplitude is unity
when mode energy equals the rotational energy of the sphere. This is
similar to setting the dimensionless amplitude to be the ratio between
displacement and radius at the surface.} We adopt
their result here. We further take $\delta\omega = 0$ and $\gamma_2 =
\gamma_1 \sim 10^{-7}$ to obtain the lowest possible threshold amplitude. For 
inertial-modes of interest, ${\xi/R_J}|_{\rm para} \sim 10^{-3}$. So
parametric damping of the tidally forced inertial-modes is
un-important.

Another three-mode coupling of consequence is between the inertial-mode, itself
and a mode at twice the frequency (up-conversion). 
However, in the case of Jupiter, twice the tidal frequency falls outside
the inertial-mode range.


Unconsidered here is another form of parametric resonance:
simultaneous excitation of two inertial-modes by the tidal potential,
with frequencies of the two modes summing up to the tidal frequency.
We find this to be also negligible for Jupiter-Io system, but likely
important for exo-jupiters.


\subsection{Comparison with \citet{gordon}}
\label{subsec:previouswork}

The most relevant work to compare our results against is that of OL,
which is an independent study that appeared while we were revising our
paper. In their work, the same physical picture as that discussed here
was considered, namely, tidal dissipation in a rotating planet. They
employed a spectral method to solve the 2-D partial differential
equations which describe fluid motion forced by the tidal potential
inside a viscous, anelastic, neutrally buoyant, polytropic fluid. This
procedure directly yields the value of the tidal torque on the planet,
without the need of a normal mode analysis. The numerical approach
allows them to include the effect of a solid core, as well as that of
a radiative envelope.  Overall, they concluded that inertial-waves can
provide an efficient mechanism for tidal dissipation, and that the
tidal $Q$ factor is an erratic function of the forcing frequency. We
concur with these major conclusions.

However, many technical differences exist between the two works. To
better understand both works, it is illuminating to discuss some of
these differences here.

Firstly, as is mentioned in \S \ref{subsec:core}, while we obtain
global inertial-modes which have well defined WKB properties and
discreet frequencies, OL demonstrated that the tidally-forced response
of a planet is concentrated into characteristic rays which are
singular lines in the limit of zero viscosity. While viscous
dissipation in their case occurs in regions harboring these rays, our
inertial-modes are predominately dissipated very near the surface (the
'singularity belt'). Moreover, although both our $Q$ values exhibit
large fluctuations as a function of tidal frequency, the origin of the
two may be different -- in our case, a deep valley indicates a good
resonance between the tide and a low-order inertial-mode, while the
situation is less clear in their case. All these differences may
originate from the presence (absence) of a solid core in their (our)
study. We are currently investigating the underlying mathematical
explanation for these differences. Again, it is interesting to note
that as the core size approaches zero, inertial-modes seem to reappear
(Ogilvie, 2004, private communication).

Secondly, OL's results are based on a $n=1$ polytrope, for which we
find that tidal coupling is vanishingly small (see Appendix
\ref{sec:polyn}),\footnote{Although we only present results for a
$\beta=1$ power-law model, they apply to a $n=1$ polytrope as well
since the two behave similarly near the surface and near the core.} 
and that inertial-modes are not important for tidal dissipation. It is
currently unclear whether this difference arises from the presence of
a core or from the presence of a radiative envelope in their
study. Despite a steep suppression of the tidal overlap integrand near
the center (integrand $\propto r^6$), the presence of a solid core may
affect tidal overlap in a more substantial manner by reflecting
inertial-waves and changing their mode structure (\S
\ref{subsec:core}).  Meanwhile, a surface boundary condition specified
at a finite density (instead of at $\rho = 0$) may cause extra tidal
coupling (\S
\ref{subsec:radiative}), as is shown by the analysis in Appendix
\ref{sec:polyn}. This issue is more relevant for extra-solar hot jupiters 
which have deeper radiative envelopes.

Thirdly, OL assumed a constant Ekman number throughout the entire
planet. Since
\be
Ek \equiv {{\gamma}\over{\omega}} = {{\nu}\over{\omega R^2}},
\label{eq:defineEk}
\ee
this implies a viscosity $\nu = \omega R^2 Ek \sim 2\times 10^{16} Ek$
that is constant throughout the planet. We have argued that the
effective viscosity value for the equilibrium tide should be of order
$\sim 10^{4} \cm^2/\s$ (\S \ref{subsubsec:gammaequi}), or an effective
$Ek \sim 10^{-13}$. However, such a weak viscosity is much smaller
than is currently reachable by a numerical method in a reasonable
amount of time. Instead, OL have opted for an alternative treatment in
which they steadily decreased the Ekman number from $Ek = 10^{-4}$ to
$10^{-7}$ and argued (based both on numerical evidence and on an
analytical toy-model) that the final $Q$ value is independent of the
Ekman number. This contrasts with our results that $Q$ roughly scales
as $\gamma_0^{-1/2}$ (\S \ref{subsec:uncertainvis}), obtained for
realistic viscosity profiles, where $\gamma_0$ is the damping rate for
a mode of wavenumber $\lambda_0$.

To make the comparison more appropriate, we adopt a constant viscosity
inside the planet and find that mode damping rates $\gamma = 5\times
10^{-9} (Ek/10^{-7})\, (\lambda/7.59)^{3.0} \s^{-1}$ in the Jupiter
model D, while individual mode $Q_{\rm res}$ value also scales
linearly with the Ekman number (eq. [\ref{eq:Qres}]). Applying
scalings derived in \S \ref{subsubsec:Qres}, we find an overall $Q
\sim 2.3\times 10^6 (Ek/10^{-7})^{-1}$.
This value is consistent with that obtained by OL for $Ek \sim
10^{-7}$.  Meanwhile, the equilibrium tide gives rise to $Q_{\rm equi}
\sim 4\times 10^6 (Ek/10^{-7})^{-1}$. So in models of a constant Ekman
number, inertial-modes contribute comparably to tidal dissipation as
does the equilibrium tide, but no better.
These results are presented in Fig.\ref{fig:Qjupiter-Dflat}.

%








\section{Summary}
\label{sec:summary}

In a series of two papers (Paper I \& this), we have examined the
physical picture of tidal dissipation via resonant
inertial-modes. This applies to a neutrally-buoyant rotating object in
which the tidal frequency in the rotating frame is less than twice the
rotation frequency.

In Paper I, we first demonstrate that under some circumstances
(power-law density profiles of the form $\rho \propto (1-r^2)^\beta$),
the partial differential equations governing inertial-modes can be
separated into two ordinary differential equations with
semi-analytical eigenfunctions. We also show that this method can be
extended to apply to more general density profiles, with the price
that the solution is exact in the surface region but only approximate
in the WKB regime. Nevertheless, this approximate solution allows us
to draw many physical conclusions concerning inertial-modes, including
their spatial characteristics, their dispersion relation, their
interaction with the tidal potential and with turbulent
convection. This semi-analytical technique gives us an edge over
current computational capabilities, though full confirmation of our
conclusions may require careful and high-resolution numerical
computation. It is clear from our study that any numerical approach
would need to be able to resolve the so-called ``singularity belt''
near the surface where inertial-modes vary sharply, and that numerical
results need to be taken cautiously when evaluating the tidal overlap.

\begin{figure}
\centerline{\psfig{figure=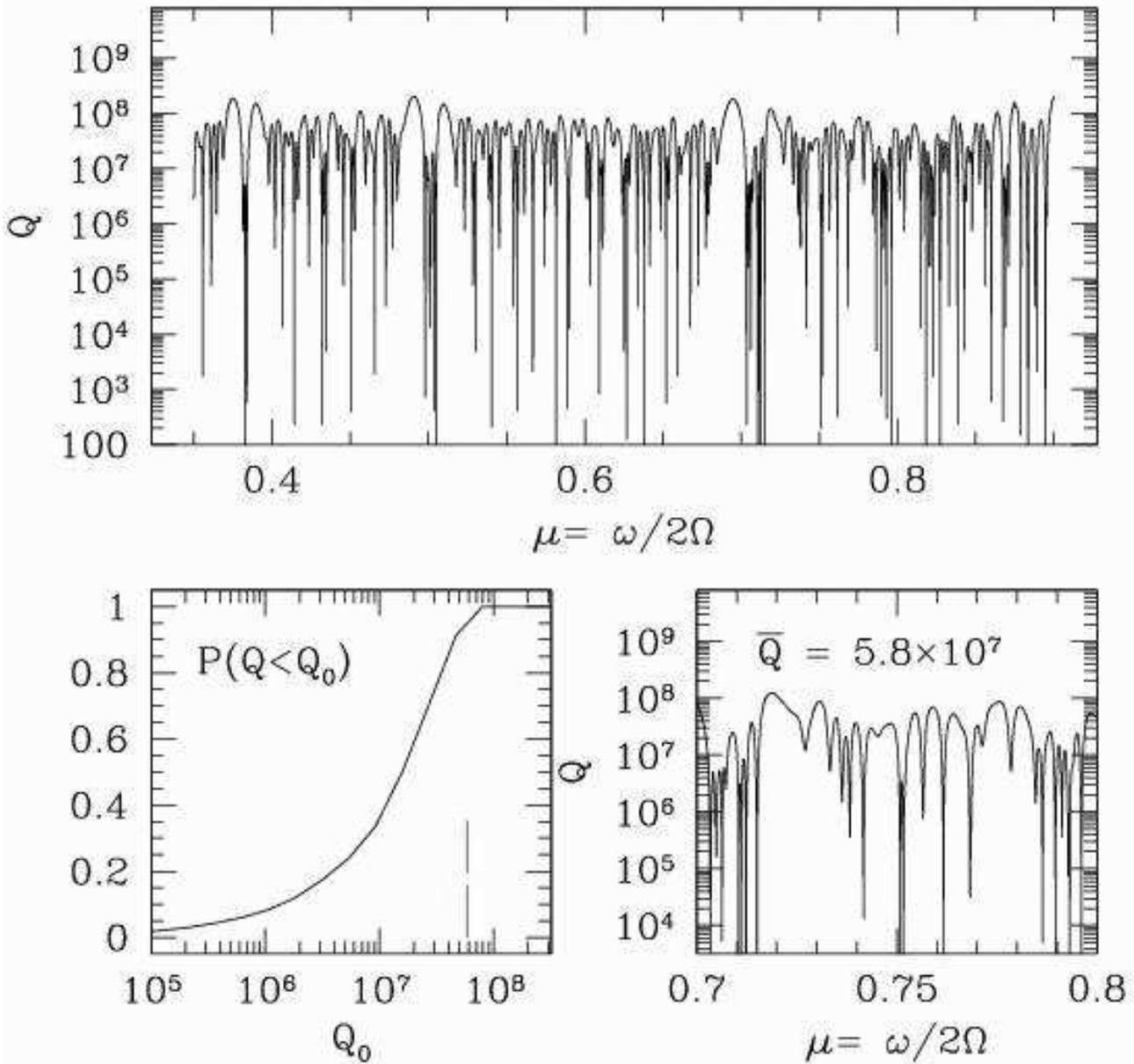,width=1.00\hsize}}
\caption{
Same as Fig. \ref{fig:Qjupiter-B}, but for model D in which the tidal
coupling decreases as $1/2n^2$ as opposed to $1/n^3$. This is related
to the presence of a first-order phase transition at $r/R \sim 0.8$.
While showing overall similar characteristics as those in
Fig. \ref{fig:Qjupiter-B}, ${\bar Q}$ has now been reduced to $\sim
5.8\times 10^7$ between $\mu = 0.7$ and $0.8$, and at any given tidal
frequency, there is a $\sim 10\%$ chance that Jupiter exhibits $Q <
10^6$.
}
\label{fig:Qjupiter-D}
\end{figure}

\begin{figure}
\centerline{\psfig{figure=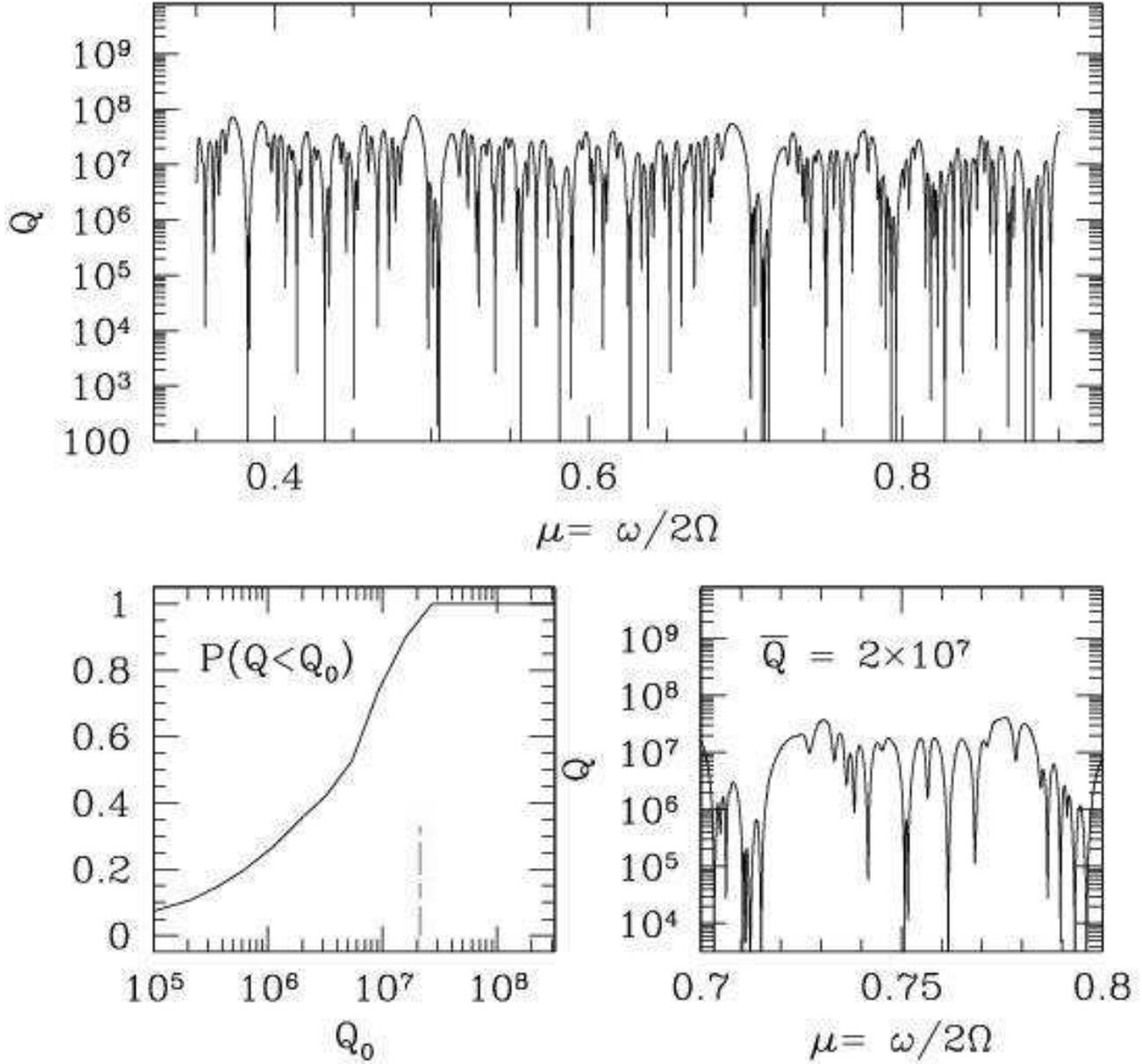,width=1.00\hsize}}
\caption{
Same as Fig. \ref{fig:Qjupiter-D}, but calculated for model D when the
index for viscosity reduction is taken to be $s = 1$
\citep{zahn} instead of $s=2$. Mode damping rates now behave as $\gamma = 10^{-10}
(\lambda/7.59)^{4.5} \s^{-1}$.  We obtain ${\bar Q} \approx 2\times
10^7$, with $\sim 30\%$ chance that $Q < 10^6$ for the current tidal
frequency. Moreover, $Q_{\rm equi} \approx 10^9$ in this case.
}
\label{fig:Qjupiter-Ds1}
\end{figure}

\begin{figure}
\centerline{\psfig{figure=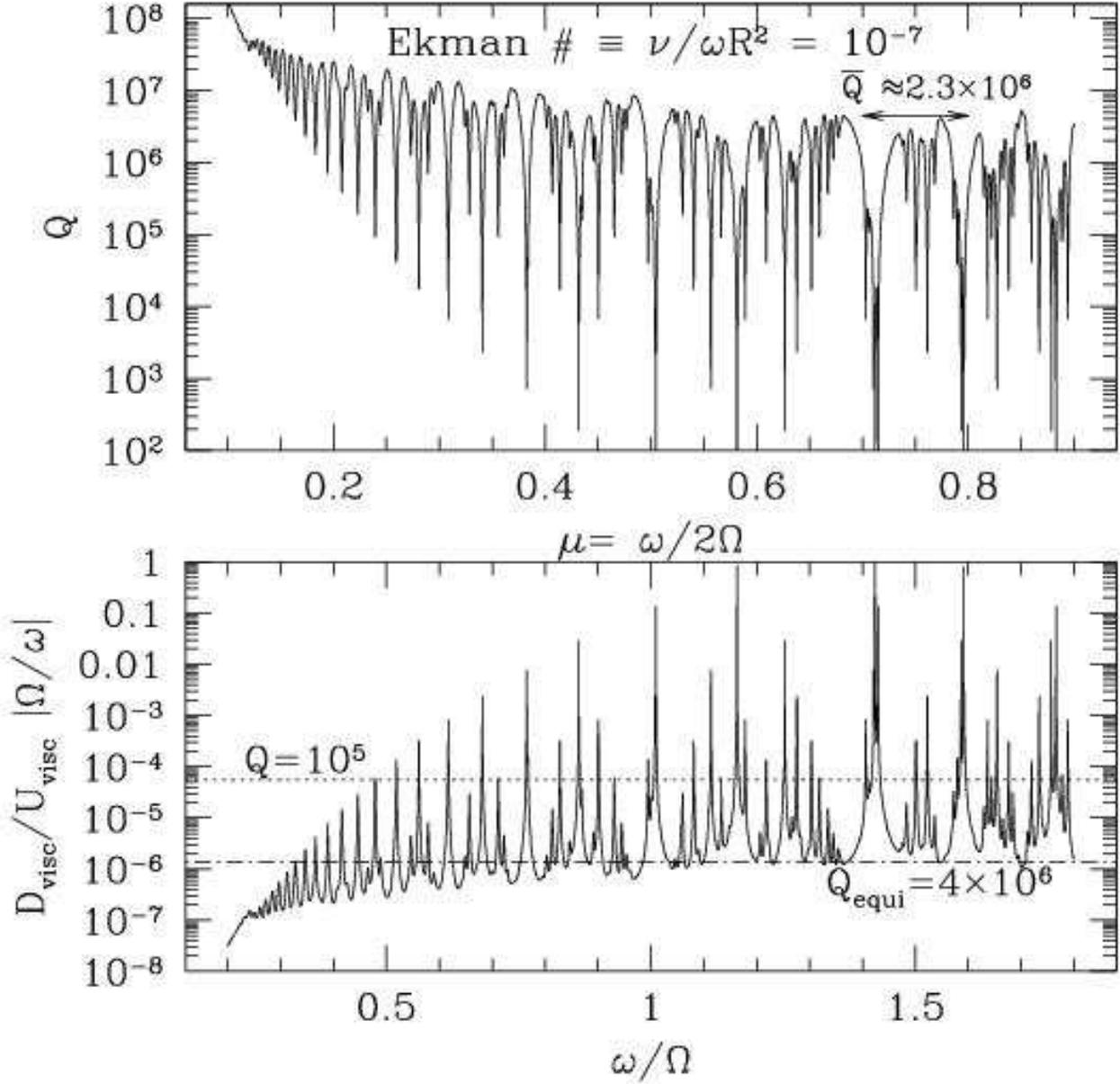,width=1.00\hsize}}
\caption{
We repeat our calculation for model D, taking the turbulent viscosity
to be a constant throughout the planet with the Ekman number $Ek =
\nu/\omega R^2 = 10^{-7}$. Mode damping rates scale much less steeply with 
inertial-mode wave-numbers, $\gamma = 5\times 10^{-9}
(\lambda/7.59)^{3.0} \s^{-1}$. The resulting $Q$ value from
inertial-modes is plotted against $\mu$ in the upper panel, with
${\bar Q} \approx 2.3\times 10^6$ over the range $\mu \in [0.7,0.8]$.
A similar calculation for model B yields ${\bar Q} \approx 3.5\times
10^6$.
We find that the ${\bar Q}$ value is inversely proportional to the
Ekman number.
%
%
The lower panel translates the $Q$ result into a quantity used in
Fig. A2 of OL (the dimensionless viscous dissipation rate, $\propto
1/Q$), plotted here as a function of $2\mu =
\omega/\Omega$.  The two overlaid lines with $Q = 10^5$ (dotted) 
and $Q = 4\times 10^6$ (dot-dashed), respectively represent the
empirically inferred $Q$ value for Jupiter and the $Q$ value
associated with the equilibrium tide in this model.
These results resemble those presented in Fig. A2 of OL for the same
Ekman number.
%
}
\label{fig:Qjupiter-Dflat}
\end{figure}

In this paper, we discuss the role in tidal dissipation played by
inertial-modes. This depends on the following three parameters: how
well coupled an inertial-mode is to the tidal potential, how strongly
dissipated an inertial-mode is by turbulent viscosity, and how densely
distributed in frequency are the inertial-modes. We have obtained all
three parameters using both toy models and realistic Jupiter
models. Low-order inertial-modes, if in resonance ($\delta\omega <
\gamma$, where $\delta \omega$ is the frequency detuning
between the tidal frequency and the mode frequency, $\gamma$ is the
mode damping rate), can dissipate tidal energy with $Q$ as small as $Q
\sim 10$. However, such a resonance is not guaranteed at all tidal 
frequencies, and the system sweeps through a fortuitously good
resonance with speed. Inertial-modes most relevant for tidal
dissipation are those satisfying $\delta \omega \sim \gamma$, where
$\delta\omega$ decreases with mode wave-number as $\delta\omega
\propto \lambda^2$, and $\gamma$ rises steeply with mode
wave-number. These are inertial-modes with wave-numbers $\lambda \sim
60$ (or total number of nodes $n = n_1 + n_2 \sim 30$). At any tidal
frequency, one can always find resonance with one such mode. They
provide the continuum to the $Q$ value, whereas previously mentioned
good resonances appear as dense valleys superposed on this continuum
(see, e.g., Fig. \ref{fig:Qjupiter-D}).

The continuum $Q$ value depends sensitively on the presence of density
discontinuities inside Jupiter, as the latter influences strongly the
magnitude of coupling between the tidal potential and inertial-modes.
Current Jupiter models show a sharp change in the adiabatic index near
the surface (hydrogen ideal-gas to Coulomb gas transition), this
warrants a $Q$ value of $\sim 10^9$. The presence of a discontinuity
in density gradient due to a phase transition (metallic hydrogen phase
transition and/or helium/hydrogen separation) has the same effect. On
the other hand, if the phase transition is first-order in nature and
incurs a density jump, $Q \sim 10^7$. Our results are uncertain up to
perhaps, one order of magnitude. But it is already clear that
inertial-modes cause much stronger dissipation than the equilibrium
tide, which yields $Q_{\rm equi} \approx 10^{12}$. In the case of
$Q\sim 10^7$, there is a $\sim 10\%$ chance that the current $Q$ value
falls between $10^5$ and $10^6$ (the empirically inferred $Q$ range for
Jupiter).

Our model also builds on the assumption that Jupiter is neutrally
stratified and turbulent all the way up to the photosphere,
as turbulent dissipation for inertial-modes with $\lambda \sim 60$ are
calculated to arise mostly near or below the photospheric
scale-height. Effects like water condensation may alter the static
stability in Jupiter's atmosphere, making the atmospheric
stratification a function of space and time.

We also restrict ourselves to core-less Jupiter models. Our conclusion
is little affected when we include an inner core with a size that is
compatible with current constraints. However, this is assuming that
global inertial-modes still exist in the presence of a solid
core. \citet{gordon2} extended the study in \citet{gordon} and
demonstrated that a new kind of tidal response appears when Jupiter
has a core: fluid motion is tightly squeezed into 'characteristic
rays' which becomes singular when the viscosity goes to zero. This is
a drastically different picture than the global eigenmode picture
described here and may lead to different $Q$ factors.

We have adopted the \citet{goldreichkeeley} prescription ($s=2$) to
account for the reduction in turbulent viscosity when the convective
turn-over time is long relative to the forcing period.  Calculations
adopting Zahn's prescription ($s=1$) produce no difference in the $Q$
value caused by inertial-modes, though we find the equilibrium tide is
significantly more strongly damped. Concerning possible effects of
nonlinearity: The surface movement of inertial-modes is predominately
horizontal. For inertial-modes that are most relevant for tidal
dissipation, the surface displacement amplitude $\sim 10^3 \cm$, or
$\sim 10^{-7}$ of the radius. We estimate that nonlinear effects are
negligible.

In our theory, tidal heat is deposited extremely close to the planet
surface (inside the 'singularity belt') and can be lost quickly to the
outside. For Jupiter, the tidal luminosity in this region is smaller
than (or at worst comparable to) the intrinsic luminosity and so would
not much alter the structure. However, there remains the intriguing
possibility that the negative angular momentum deposited to the belt
may affect surface meteorology (jet streams and
anticyclones). Moreover, if this theory also applies to hot
exo-jupiters, the tidal luminosity is unlikely to be responsible for
inflating planets and solving the size-problem of close-in exo-jupiter
HD209458b.

Although our investigation was stimulated by the fact that exo-solar
planets exhibit similar $Q$ values as Jupiter does, it may be
difficult to draw a close analogy between Jupiter and hot
exo-jupiters: the existence of a first-order phase transition is less
convincing in the latter due to their hotter interiors; the upper
atmosphere of these planets are strongly irradiated by their host
stars and are therefore likely to be radiative; they may have rather
different core sizes depending on their formation
history. Nevertheless, it is our plan to extend the current study to
exo-jupiters, as investigations into these bodies may ultimately yield
clue for the story of Jupiter. It is also foreseeable that the theory
developed here has implications for Saturn, Uranus, solar-type
binaries, M-dwarfs and brown-dwarfs.



\begin{acknowledgements}
Phil Arras has contributed to the early stages of this work. I thank
him for an enjoyable collaboration. I also acknowledge stimulating
conversations with Gordon Ogilvie and Doug Lin, and thank Tristan
Guillot for making his Jupiter models publicly available.  Lastly,
this article benefited greatly from the insightful comments by the
referee, David Stevenson.
\end{acknowledgements}


\begin{appendix}

\section{Relevant Properties in a Jupiter Model}
\label{sec:jupiter}

Here, we study properties of Jupiter that are relevant for the tidal
process. This is based on publicly available models of Jupiter
presented in \citet{guillotreview}. They are produced with the newest
equation of state and opacity calculations, including the effect of
hydrogen phase transition, and alkali metal opacity. They satisfy
gravity measurements (esp. $J_2$ \& $J_4$) to much better than a
percent and reproduce other global properties of Jupiter (radius,
surface temperature, intrinsic flux).

In this study, we focus on two models, B and D, out of the five sample
models presented in \citet{guillotreview}. Model B is produced with an
interpolated hydrogen equation of state (meaning no first-order
metallic hydrogen phase transition), and has no heavy metal
core. Model D, in contrast, contains a first-order phase transition
(PPT equation of state) and has a core with mass $10 M_\oplus$. The
photosphere for both these models is located at a radius of $\approx
7\times 10^9 \cm$, at a pressure of $\approx 10^6
\dyne/\cm^2$, and with a temperature $170 \K$ and a density 
$1.6\times 10^{-4} \g/\cm^3$.


The interiors of these models are fully convective (outside the
core). Due to the high density in Jupiter (mean density $\sim 1.3
\g/\cm^3$), the convection speed needed to carry the small intrinsic flux  
($ 5.4\times 10^3 \erg/\s/\cm^2$) is highly subsonic, resulting in an
almost exactly adiabatic temperature profile (super-adiabatic gradient
$\sim 10^{-8}$ or smaller). This justifies our assumption of neutrally
buoyant fluid when investigating inertial-modes. Only the thin
atmosphere above the photosphere, with a local pressure scale height
$\sim 20\km$, is radiative.

\subsection{Density Profile}
\label{subsec:density}

Two features in the density profile of these models deserve attention.

At radius $r/R \sim 0.8$, pressure $\sim 10^{12} \dyne/\cm^2$, and
density $\sim 1 \g/\cm^3$, hydrogen undergoes a phase
transition. Above this layer, hydrogen is mostly neutral and
molecular. Below this layer, the mean atomic spacing becomes smaller than
a Bohr radius and electrons are pressure ionized. The strong Coulomb
interaction and electron degeneracy resemble those in a metal and the
transition is referred to as 'liquid metallic hydrogen' transition
\citep{guillotreview}. The nature of this transition is
still poorly understood. Model B assumes this transition is of
second-order and entails a discontinuity only in the gradient of
density (of order $50\%$), while model D assumes it is a first-order
transition with a density jump of order $10\%$. These two different
treatments should bracket the actual equation of state of hydrogen.

Another feature sets in nearer the surface, at radius $r/R \sim 0.98$,
pressure $\sim 10^{10} \dyne/\cm^2$ and density $0.1
\g/\cm^3$. Above this region, the gas can be considered
as ideal diatomic gas ($H_2$). As the temperature is below
$2000 \K$, the mean degree of freedom for each molecule is $5$ (three
translational plus two rotational).\footnote{This number is smaller
near the photosphere when the temperature cools toward the rotational
temperature of $H_2$ ($85\K$). Not all rotational levels are populated
\citep{SCVH}.} The specific heat per
molecule at constant volume and constant pressure are, respectively,
$C_V = 5/2 k_B$, $C_p = 7/2 k_B $, yielding $\Gamma_1 = {\partial \ln
P/\partial \ln \rho}|_s = C_p/C_V= 1.4$. Below this region, however,
$\Gamma_1$ rises to $\sim 1.8-2.2$ in the main body of the planet, and
approaches $ 3$ in the very deep interior
\citep{stevenson78,stevensonreview}.

This results in different density profiles above and below this
region.  Recall our definition of $\beta$: $\rho \propto
[1-(r/R)^2]^\beta$. The Jupiter models show that $\beta \sim 1.8$
(corresponding to $\Gamma_1 \sim 1.4$) above this layer, while $\beta
\sim 1$ (corresponding to $\Gamma_1 \sim 2$) in the interior.  We also
observe that this transition of $\beta$ occurs over a fairly narrow
region of radial extent $\Delta r \sim 0.02 R$, or $\sim 4$ local
pressure scale height. As is discussed in \S
\ref{subsubsec:overlapbeta}, this transition is of significance to our
tidal coupling scenario.

\begin{figure}
\centerline{\psfig{figure=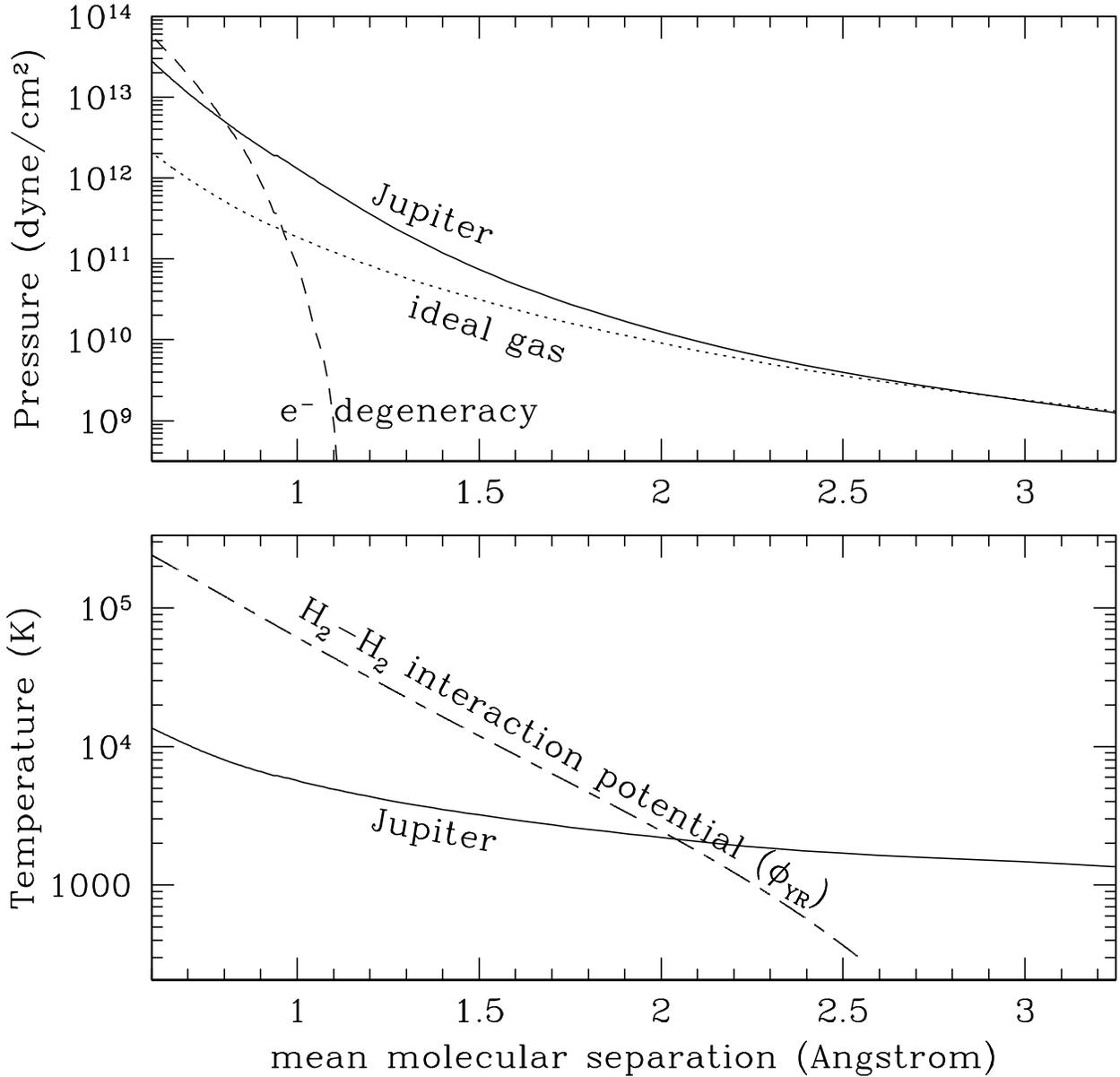,width=1.00\hsize}}
\caption{
The effect of $H_2$-$H_2$ interaction on gas pressure. The upper panel
shows pressure as a function of mean molecular separation (in \AA),
with lower density to the right. The solid line is the actual Jupiter
profile in a model from
\citet{guillotreview}. The dotted line represents the ideal gas contribution
(thermal pressure) -- it falls short of explaining the total pressure
above a pressure $p \sim 10^{10} \dyne/\cm^2$. The dashed curve shows
the contribution from electron degeneracy which only becomes important
for $p \geq 10^{12}\dyne/\cm^2$. For the in-between region, another
pressure contribution has to kick in.  The lower plot examines what
this extra contribution is. Here, gas temperature inside Jupiter is
plotted as a function of the mean separation (solid curve), while the
dashed curve depicts the inter-particle potential in unit of Kelvin
\citep[the $\phi_{YR}$ potential from][]{rry}. Molecular interaction
is repulsive for a separation below $\sim 3$ \AA and the interaction
energy becomes comparable to the thermal energy at a separation $\sim
2$ \AA.  This contributes to the gas pressure. As density rises, the
increasingly repulsive interaction dominates the gas pressure and
causes the pressure to rise with density more steeply than that of an ideal
gas.}
\label{fig:H2interaction} 
\end{figure}

But what is the cause behind the rise of $\Gamma_1$ near $p \sim
10^{10} \dyne/\cm^2$? The ionization fraction of electron is too low
($\sim 10^{-6}$) in this region to make a difference by degeneracy
pressure; hydrogen is bound into $H_2$ and only starts to be
dissociated near $p \sim 10^{12} \dyne/\cm^2$. The true cause, it
turns out, is the non-ideal behavior of molecules, a little-talked
about effect. At a density of $0.1 \g/\cm^3$, the mean molecular spacing
is $\sim 2$ \AA. While the interaction potential between $H_2$ and $H_2$
molecules is mildly attractive at spacing $> 3$ \AA  (the van der Waals
force), it rises exponentially inward. By the time the spacing
decreases to below $\sim 2$ \AA, this potential is more positive than
$k_B T$ and the gas pressure is no longer dominated by thermal
pressure, but is dominated by the repulsive interaction between
molecules. This is illustrated in Fig. \ref{fig:H2interaction}.
As density rises, molecules increasingly resemble hard spheres,
leading to a steeper dependence of pressure on density, or $\Gamma_1
\sim 2$ ($\beta \sim 1$). This non-ideal effect loses out 
at $p \sim 10^{12} \dyne/\cm^2$ above which $H_2$ molecules are
dissociated and electrons are pressure ionized (the metallic hydrogen
phase).
s
%
%

\subsection{Turbulent Viscosity Profile}
\label{subsec:viscosity}

Inside Jupiter, molecular viscosity is too weak to cause any
discernible dissipation on the inertial-modes. We turn to turbulent
viscosity.

The kinematic shear viscosity is estimated from the mixing length
theory as \citep{goldreichkeeley,zahn,terquem}
\be
\nu_T \sim v_{\rm cv} \ell_{\rm cv} {1\over{1+(\omega \tau_{\rm cv}/2\pi)^s}},
\label{eq:nucv}
\ee
where $v_{\rm cv}$, $\ell_{\rm cv}$ and $\tau_{\rm cv}$ are
characteristic convection velocity, scale length and turn-over
time. The exponent $s$ describes the reduction in efficiency when
convection is slow compared to the tidal period ($\omega \,
\tau_{\rm cv} \gg 1$). Its value is still under debate, but simple
physical arguments \citep{goldreichphil,goodmanoh} have suggested that
$s=2$, while \citet{zahn} advocated for a less severe reduction with
$s=1$. We adopt $s=2$ in our main study but discuss the scenario when
$s=1$. Some previous studies have adopted a form without $2\pi$ in the
above expression. The viscosity is effectively smaller but we will
show that this does not affect the final $Q$-value significantly.

\begin{figure}
\centerline{\psfig{figure=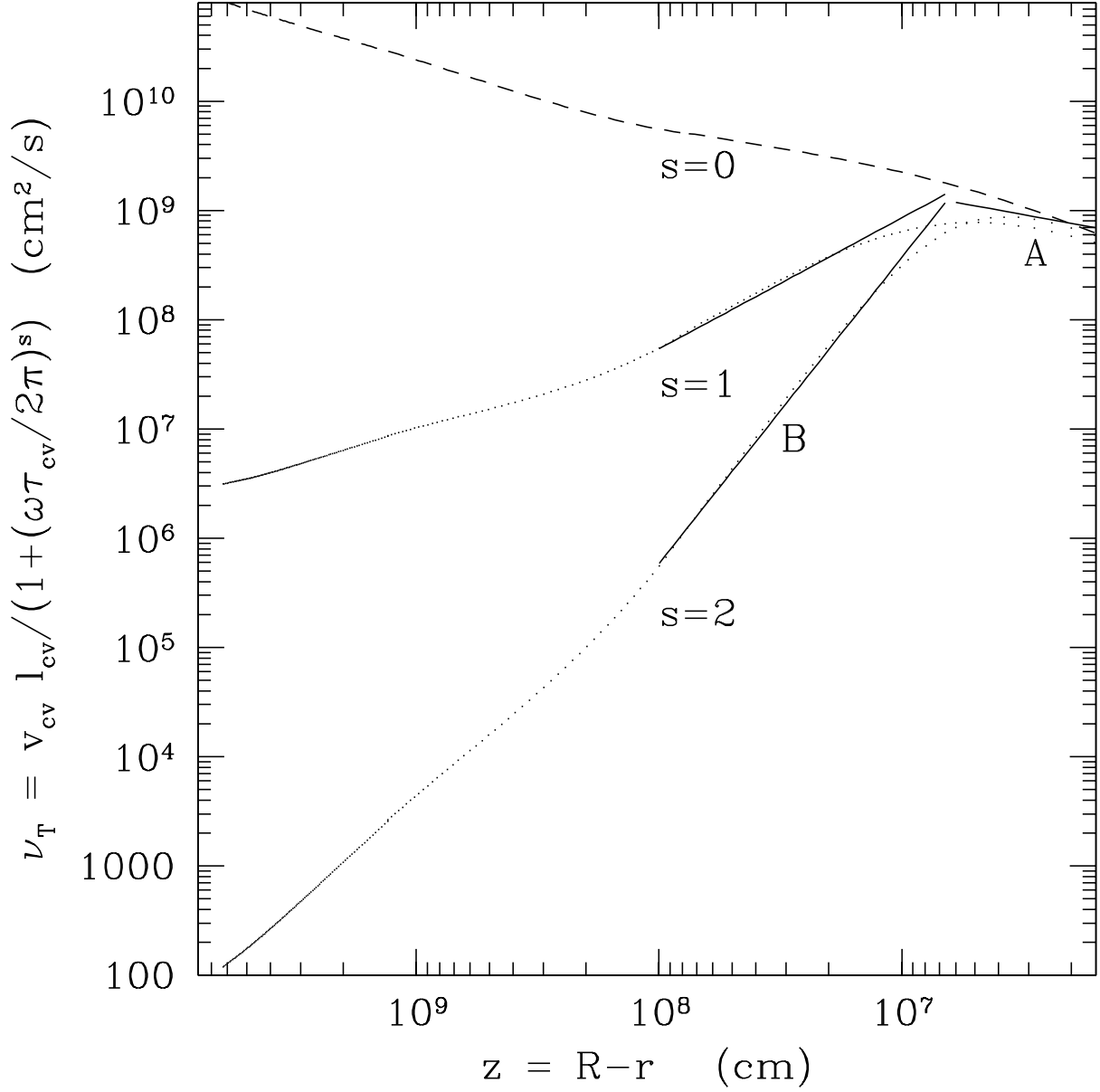,width=1.00\hsize}}
\caption{
The effective turbulent viscosity $\nu_T$ is plotted here as fine dots
against depth ($z$) for the Jupiter model, when various $s$ values are
adopted.  The dashed curve ($s=0$) is the un-reduced turbulent
viscosity (corresponding to $s=0$). The reduced viscosity (dotted
curves) deviate from this curve below a depth $z_{\rm crit} \sim
10^{-2.8} R \sim 10^7 \cm$ at which $\omega
\tau_{\rm cv}/2\pi \sim 1$. 
Above $z_{\rm crit}$, the viscosity is well described by line A:
$\nu_T
\sim 2\times 10^{10} (z/R)^{1-\beta/3} \propto z^{0.4}$ (with $\beta=1.8$ 
in the model).  Below this depth, reduction is important and $\nu_T
\sim 4 (z/R)^{-1-\beta} \propto z^{-2.8}$ for $s=2$ (straight line B)
and $\nu_T \sim 3 \times 10^5 (z/R)^{-2\beta/3} \propto z^{-1.2}$ for
$s=1$. Deeper down ($z > 10^8\cm$), as $\beta$ value is varied from
$1.8$ to $1$, $\nu_T$ takes on a different scaling with
depth. However, this is irrelevant as turbulent dissipation from the
deep interior is insignificant.}
\label{fig:guillot-viscosity}
\end{figure}

In mixing length theory, $v_{\rm cv} \approx ({\rm flux}/\rho)^{1/3}$,
$\tau_{\rm cv}\approx \ell_{\rm cv} /v_{\rm cv}$, and $\ell_{\rm cv}
\approx H \approx z/\beta$,  where $H$ is the density scale height, $z$ 
is the physical depth ($z = R-r$), and $\beta$ appears in the density
power-law as $\rho = [1-(r/R)^2]^\beta \propto z^\beta$ for $z \ll
R$. Let the depth at which $\omega \tau_{\rm cv}/2\pi \approx 1$ be
$z_{\rm crit}$. Above $z_{\rm crit}$, $\nu_T$ depends on $z$ weakly,
\be
\nu_T \propto z^{1-\beta/3},
\label{eq:nu2}
\ee
while below this layer, the turbulent viscosity is significantly
reduced and $\nu_T$ decreases sharply inward as
\be
\nu_T \propto z^{-1-\beta},
\label{eq:nu3}
\ee
when $s=2$ and 
\be
\nu_T \propto z^{-2\beta/3},
\label{eq:nu4}
\ee
when $s=1$. These approximate scalings are shown in
Fig. \ref{fig:guillot-viscosity} for Jupiter models B \& D. They
compare well with numerical results.

\section{Tidal Overlap in a Constant Density Sphere}
\label{sec:poly0}

In a constant density sphere, $m=-2$ inertial-modes are expressed in
the following form (Paper I)
\be
\rho^\prime = {{\omega^2 \rho^2}\over{\Gamma_1 p}} \psi = 
A R^2 \, {{\omega^2 \rho^2}\over{\Gamma_1 p}} P_{\ell}^{-2} (x_1)\,
P_{\ell}^{-2} (x_2),
\label{eq:recap}
\ee
where $A \ll 1$ stands for the dimensionless amplitude of $\psi$, and
$R$ is the radius of Jupiter. This density perturbation is related to
the equilibrium tide value $\rho^\prime_{\rm equi}$ as
\be
{{\rho^\prime}\over{\rho^\prime_{\rm equi}}} = {2\over 3}
{{\omega^2}\over{\omega_{\rm Io}^2}}\,{{M_J}\over{M_{\rm Io}}} \, A\,
(1-\mu^2) g_1(x_1) g_2(x_2),
\label{eq:ratiorhoprime}
\ee
where Io's orbital frequency $\omega_{\rm Io} = (G M_{\rm
J}/a^3)^{1/2}$, $M_J$ and $M_{\rm Io}$ are the masses of Jupiter and
Io, and the dimensionless frequency $\mu = \omega/2\Omega = 0.766$.
The function $g_i(x_i) = P_\ell^{m}(x_i)/(1-x_i^2)$ (introduced in
Paper I).

Pressure in a constant density ($\rho = \rho_0 = {\rm const}$),
self-gravitating sphere is given by $p = p_0 [1-(r/R)^2]$ where $p_0 =
2\pi/3\, G R^2 \rho_0^2 = 3/(8\pi) GM^2/R^4 $ with $M$ being the total
mass. Since $[1-(r/R)^2] = (x_1^2 - \mu^2)(\mu^2 -
x_2^2)/(1-\mu^2)/\mu^2$, and volume elements in Cartesian coordinates
and ellipsoidal coordinates are related to each other as $dxdydz =
(x_1^2 - x_2^2)/(1-\mu^2)/\mu\, dx_1 dx_2 d\phi$, we obtain the
following tidal overlap,
\be
- 
\int \delta\Phi_{\rm tide} \rho^\prime d^3 r = {{9}\over 4}
{{\omega^2 R^5 M_{\rm Io}}\over{\Gamma_1 a^3}}\, A
\, \int_{-\mu}^\mu \int_\mu^1 {{\mu}\over{(1-\mu^2)}}
{{(1-x_1^2) (1-x_2^2) (x_1^2 - x_2^2)}\over{(x_1^2 - \mu^2) (\mu^2 -
x_2^2)}}\, P_\ell^{-2} (x_1) P_\ell^{-2} (x_2)\, dx_1 dx_2.
\label{eq:overlappoly0}
\ee
The spatial integration can be symbolically performed by Mathematica
(best done after conversion to spherical coordinates) and it yields
$\sim 0.4 (1-\mu^2)/\lambda$ where $\lambda^2 = \ell(\ell+1) -
|m|(|m|+1)$.
%
%
%
So the overlap is 
\be
- 
\int \delta\Phi_{\rm tide} \rho^\prime d^3 r = 0.4 \, {{9}\over 4}
{{\omega^2 R^5 M_{\rm Io}}\over{\Gamma_1 a^3}}\, A\, {{(1-\mu^2)}\over
\lambda}.  
\label{eq:repeat}
\ee
However, the constant density case is pathological: the value of
$\Gamma_1$ formally approaches infinity for incompressible
fluid. Inertial-modes could not cause any density fluctuation (eq.
[\ref{eq:recap}]) and the tidal overlap is formally zero.\footnote{The
equilibrium tide, on the other hand, has finite tidal overlap. It is
equivalent to an inertial-mode with $\ell = 2$ so its spatial overlap
diverges near the surface as $p$ approaches $0$, counteracting the
formally infinite $\Gamma_1$.}

If we take $p = constant$ over the entire sphere (so $\Gamma_1$ is a
finite constant), only two motion have non-zero overlap with the tidal
potential: the equilibrium tide and the two lowest order even-parity
inertial-modes with $\ell =4$. This fact has been pointed out in
\citet{papaloizousavonije} when they
considered the convective core of early-type stars.

\section{Tidal Overlap in a Single Power-law Model}
\label{sec:polyn}

Are inertial-modes in power-law models coupled to the tidal potential?

In paper I, we show that one can obtain exact solutions for
inertial-modes when the density profile is a single power-law $ \rho
\propto [1-(r/R^2)]^\beta$. This allows us to show that inertial-modes 
in single power-law models do not couple appreciably to the tidal
potential, except for the two lowest order even-parity modes
(corresponding to the $\ell =4$ modes in the constant density
case).\footnote{If we adopt conventional polytrope models with $p
\propto \rho^{1+1/\beta}$, we can obtain  approximate solution for 
the inertial-modes (Paper I). We find that they give essentially the
same tidal overlap results as single power-law models of the same
$\beta$.} Moreover, the coupling strength falls off with increasing mode
order as a power-law with the index related to the polytrope index.

The angular dependence of each even-parity, $m=-2$ inertial-mode can
be decomposed into
\be
\psi = 
\psi_1 (x_1) \psi_2 (x_2) = 
\sum_{\ell = 2}^{\infty} P_\ell^{-2} (\theta) C_\ell(r),
\label{eq:decomp}
\ee
where
\be
C_\ell (r) = \int \psi_1 (x_1) \psi_2 (x_2) P_\ell^{-2} (\theta) \sin
\theta d\theta
\label{eq:Cl}
\ee
is non-zero for $\ell = 2$. 
$C_\ell(r)$ is an oscillating function of the radius $r$. We find
numerically that $C_2(r) \propto r^2$ near the center, while near the
surface $C_2(r)$ approaches a constant for $\beta > 1$, and $\propto
[1-(r/R)]^{(\beta-1)/(\beta+1)}$ for $0 < \beta < 1$. 
The tidal overlap integral is reduced to the following radial
integral,
\be
- \pi \int
\delta\Phi_{\rm tide} {{\omega^2 \rho^2}\over{\Gamma_1 p}} \psi d^3 r
=  \pi \sqrt{{32\pi}\over{15}}\, {{3GM_{\rm Io}}\over{2a^3}}\omega^2 
\int_0^R D_2(r) \, dr =  \pi \sqrt{{32\pi}\over{15}}\, 
{{3GM_{\rm Io}}\over{2a^3}}\omega^2
\int_0^R\, 
C_2 (r) 
{\rho^2\over{\Gamma_1 p}}\, r^4 dr.
\label{eq:reduction}
\ee
where we have introduced the integrand $D_2(r) = r^4 C_r (r)
\rho^2/\Gamma_1 p$. It is also an oscillating function of $r$ 
with an envelope that scales as $r^6$ near the center, and scales near
the surface as $[1-(r/R)]^{\beta-1}$ for $\beta \geq 1$, and as
$[1-(r/R)]^{\beta-{2/(\beta+1)}}$ for $0 <
\beta < 1$.
So this integral diverges near the surface if $\beta < \sqrt{2}-1$.

We find that the integral decreases with increasing mode order in a
power-law fashion with the index depending on $\beta$. In
the following, we explain the observed fall-off with a simple
toy-model.

We approximate $D_2(r)$ as a product of a rapidly oscillating function
and a slowly varying envelope. A rather accurate form turns out to be
\be
D_2(r) dr = \cos (n \Theta) f(\Theta) d\Theta, \label{eq:D2rTheta}
\ee
where the new variable $\Theta = \cos^{-1} r/R$, $n$ is an integer and
is the number of radial nodes in $D_2(r)$. We find $n = n_1 + n_2$ for
the inertial-modes.  The smooth function $f(\Theta)$ has a {\it
leading}
term of $(\pi/2 - \Theta)^6$ near the center ($\Theta \sim \pi/2$) and
a {\it leading} term of $\Theta^{2\beta-1}$ near the surface ($\Theta
\sim 0$).\footnote{Here, we focus only on models with $\beta > 1$}.
For the moment we assume $2\beta$ is an integer, and that terms of
order $\Theta^{2\beta}$ and higher also exist near the surface.

\begin{figure}
      \centering{
      \vbox{\psfig{figure=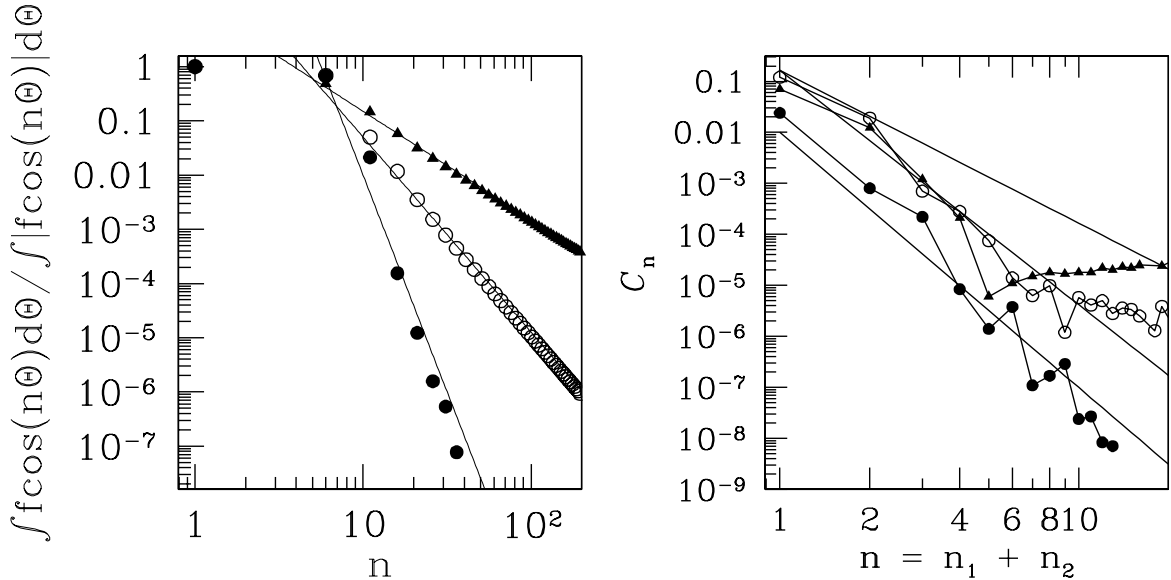,width=1.00\hsize,%
       bbllx=0pt,bblly=160pt,bburx=650pt,bbury=450pt,clip=}}
       }
\caption{
Severity of cancellation in the overlap integral as a function of mode
nodal number ($n$ where $n$ is even) in three single power-law models
(solid triangles for $\beta = 1.0$ solid circles for $1.5$ and open
circles for $1.8$).  The left-hand panel is the toy model result where
we have taken the envelope of the cosine function
(eq. [\ref{eq:D2rTheta}]) to be $f(\Theta) = f = r^6 \rho^2/p
dr/d\Theta$ where $\Theta = \cos^{-1} r/R$. This allows the toy model
tidal integrand to have the correct asymptotic behavior as the
realistic tidal integrand both near the center and near the surface.
The severity of cancellation is measured here by $\int f
\cos(n\Theta) d\Theta/\int f |\cos (n\Theta)| d\Theta$ and it scales as 
(solid lines) $n^{-2}$, $n^{-8}$ and $n^{-3.6}$, respectively, for the
three models, consistent with results in equation
\refnew{eq:evenodd}. The right panel is the severity of cancellation 
${\cal C}_n$ (eq. [\ref{eq:definecalc}]) calculated for inertial-modes
in the same three models. Again, the three straight lines are the
analytically expected scalings, $n^{-3}$, $n^{-5}$ and $n^{-4.6}$,
respectively, for the three models. The extra power of $n$ compared to
those for the toy model arises from cancellation in the angular
direction, except for the $\beta=1.5$ model, which does not fall off
as $n^{-9}$ due to the presence of $\Theta^{3}$ term near the surface.
Results in the $\beta=1.0$ model first deviates from the scaling but
returns to it at large $n$ and the $\beta=1.5$ model falls off more
steeply than the $\beta=1.8$ model, as is expected.
}
\label{fig:singlebeta} 
\end{figure}

Integrating-by-part yields
\begin{eqnarray}
\int_0^R D_2(r) \, dr & =  & 
- \int_0^{\pi/2} \cos(n\Theta) f(\Theta) d\Theta
\nonumber \\
& = & - \left.{{\sin(n\Theta) f(\Theta)}\over n}\right|_{0}^{\pi/2} 
- \left.{{\cos(n\Theta) f'(\Theta)}\over{n^2}}\right|_0^{\pi/2} 
+ \left.{{\sin(n\Theta) f''(\Theta)}\over{n^3}}\right|_0^{\pi/2}
+ \left.{{\cos(n\Theta) f'''(\Theta)}\over{n^4}}\right|_0^{\pi/2} 
+ {\cal O}\left({1\over {n^5}}\right).
\label{eq:oscillate}
\end{eqnarray}
So the value of this integral depends only on behavior of the
function $f(\Theta)$ at the two boundaries. When $n$ is an even
integer, only odd-order derivatives enter the above expression and we
obtain the following results for the tidal integral,
\begin{eqnarray}
\int_0^R D_2(r)\, dr & \approx & 
{{f^{(2\beta-1+Mod[2\beta,2])}\left(0\right)}\over{n^{2\beta+Mod[2\beta,2]}}}, 
\hskip1.0in {\mbox {if $2\beta \leq 7$}} 
\nonumber \\
& \approx & {{f^{(7)}
\left({{\pi}\over 2}\right)}\over{n^{8}}}, 
\hskip1.9in {\mbox {if $2\beta \geq 7$}}
\label{eq:evenodd}
\end{eqnarray}
where $f^{(2\beta-1)}(0) =
d^{2\beta-1}f/d\Theta^{2\beta-1}|_{\Theta=0}$ and so on. When $2\beta$
is odd, the above scaling depends on the fact that near the surface,
terms scaled as $\Theta^{2\beta}$ and higher also exist. If they do
not (as in the left panel of Fig. \ref{fig:singlebeta}), $1/n^8$
scaling prevails.

When $n$ is an odd integer, slightly different scalings apply:
\begin{eqnarray}
\int_0^R D_2(r)\, dr & \approx & 
{{f^{(2\beta-1+Mod[2\beta-1,2])}\left(0\right)}\over{n^{2\beta+Mod[2\beta-1,2]}}}, 
\hskip1.0in {\mbox {if $2\beta \leq 7$}} 
\nonumber \\
& \approx & {{f^{(6)}
\left({{\pi}\over 2}\right)}\over{n^{7}}}, 
\hskip1.9in {\mbox {if $2\beta \geq 7$}}
\label{eq:evenodd2}
\end{eqnarray}
We have confirmed these scalings numerically with a range of
expressions for $f(\Theta)$. The result only depend on the boundary
behavior of $f(\Theta)$ as long as it is sufficiently
smooth.\footnote{In Appendix \ref{sec:polyn2}, we discuss what the
meaning of 'sufficiently smooth' is.} This explains why models with
different polytrope representations ($\rho \propto [1-(r/R)]^\beta$ or
$p\propto
\rho^{1+1/\beta}$) give rise to essentially the same overlap integrals.
Moreover, when $\beta$ is a fractional number (other than an integer
or a half-integer), we find numerically that $\int_0^R D_2(r) dr
\propto 1/n^{2\beta}$ for $2\beta \leq 7$, similar to the above expressions.

Recall that the angular integration to yield $C_2(r)$ already involves
a cancellation of order $1/n$.\footnote{This is so because the
functional value at one of the two boundaries (the equator) is not
zero -- see Eq. \refnew{eq:oscillate}.} Moreover, even-parity modes
implies $n = n_1 + n_2$ to be an even number.
So for the following three power-law models, $\beta=1.0$, $\beta=1.5$
and $\beta=1.8$, we expect that the overall tidal overlap falls off
with $n$ as $n^{-3}$, $n^{-5}$ and $n^{-4.6}$, respectively. These
analytical expectations are plotted in Fig. \ref{fig:singlebeta} along
with numerical results. The agreement is reasonable, both when
integrating using the toy model ($f(\Theta) \cos(n\Theta)$) and when
integrating using realistic inertial-mode eigenfunctions.

%

In obtaining results like those presented in
Fig. \ref{fig:singlebeta}, one needs to be extremely careful with
numerical precision. Round-off errors in the numerically produced
power-law models as well as in the inertial-mode eigenfunctions may
occult the fine cancellation and lead to artificially large coupling.

\section{Tidal Overlap in Other Models}
\label{sec:polyn2}

The derivation leading to equation \refnew{eq:oscillate} assumes that
the integrand $f(\Theta)$ is sufficiently smooth. What is
'sufficiently smooth' and in what situation does this assumption break
down? It turns out that the break-down occurs for realistic planet
models and that the tidal overlap is much larger than what one obtains for
single power-law models.

The smoothness assumption is violated if $f(\Theta)$ has a discreet
jump inside the planet. Such a discontinuity is caused by the density
discontinuity associated with a first-order phase transition region
(e.g., gas-to-metallic hydrogen phase transition region at $r/R
\sim 0.80$).  Let the jump be $\Delta f$ at $\Theta = \Theta_0$.  
It contributes a term, ${{\Delta f \sin(n\Theta_0)}/ n} \sim \Delta
f/n$, to the tidal overlap. Even if $\Delta f$ is small, this term may
dominate for high order modes. Similar reasoning applies if
$f(\Theta)$ exhibits a discontinuity at a higher order derivative, for
instance, if the above mentioned phase transition is of second order
in nature so that a discontinuity in the gradient of density
exists. In this case, the contribution to the overlap integral is of
order $\sim \Delta f'/n^2$. 

The smoothness assumption can also be violated if $f(\Theta)$ is
infinitely continuous yet it (or one of its derivatives) has a sharp
transition over a small region, namely, if this transition occurs over
a width of $\Delta \Theta$ which encompasses only one node or less
($\Delta n \sim n\, \Delta \Theta/\pi/2 \leq 1$). This can be caused
by, e.g., a relatively sharp power-law index change inside the
planet. As is discussed in \S \ref{sec:jupiter}, gas pressure inside
Jupiter changes its nature from that of an ideal gas to that of
strongly interacting molecules around $r/R \sim 0.98$. Here we
observe a variation in the polytropic index over one pressure scale
height, or over a thickness of $\Delta r/R \sim 0.002$. Within this
narrow region, $f^{'}(\Theta)$ varies rapidly for an amount $\Delta
f^{'}$, and $f^{(2)}(\Theta)$ has a peak value of $\sim \Delta
f^{'}/\Delta \Theta$. The overlap integral
\begin{equation}
\int_0^{\pi/2} \cos(n\Theta) f(\Theta) d\Theta
 =  \left.{{\sin(n\Theta) f(\Theta)}\over n}\right|_{0}^{\pi/2}
+\left.{{\cos(n\Theta) f'(\Theta)}\over{n^2}}\right|_0^{\pi/2} -
{{1\over n^2}} \int_0^{\pi/2} \cos(n\Theta) f^{''}(\Theta) d\Theta,
\label{eq:oscillate2}
\end{equation}
can be dominated by the last term and yields $\Delta f^{'}/n^2$ if
$\Delta n \sim n \Delta r/R \sim n \Delta \Theta/\pi/2 \leq 1$, or $n
\leq 1/0.002 \sim 500$. For $n \geq 500$, $f(\Theta)$ can be 
considered as sufficiently smooth and the analysis in Appendix
\ref{sec:polyn} applies.


We numerically confirm these conclusions by integrating $f(\Theta)
\cos(n\Theta)$ using a range of density profiles. Here, we take $f(\Theta) = 
r^6 \sqrt{\rho_{\rm surf}/\rho} \rho^2/p\, dr/d\Theta$, where
$\rho_{\rm surf}$ is $[1-(r/R)^2]^{\beta}$ with the $\beta$ value
taken at the surface. This $f(\Theta)$ has the same asymptotic
behavior as $D_2(r)$ near both boundaries.

\begin{figure}
\centerline{\psfig{figure=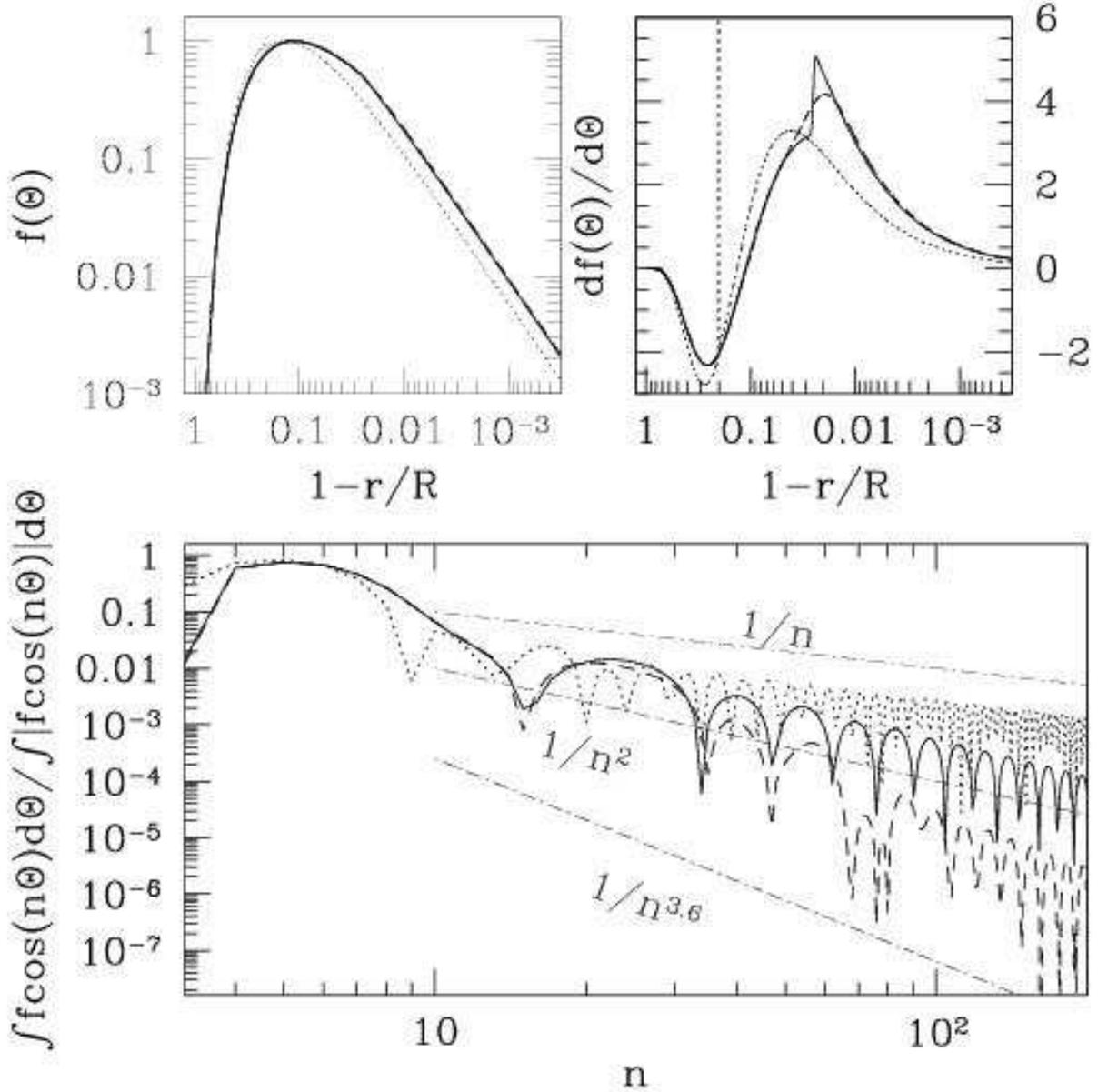,width=1.00\hsize}}
\caption{
The severity of cancellation in the overlap integral calculated using
the toy model for three different density profiles are shown in the
lower panel as a function of $n$ ($n$ even), while the top two panels
show the corresponding $f(\Theta)$ (left) and $df/d\Theta$ (right) as
functions of $1-r/R = 1- \cos\Theta$.  We take $f(\Theta)$, the
envelope of the cosine function to be $f(\Theta) = r^6 \rho^2/p
(\rho_{\rm surf}/\rho)^{1/2} dr/d\Theta$ while the various density
profiles are: a $\beta=1.8$ power-law model, overlaid with a $1\%$
density jump at $r/R = 0.8$ (dots, exhibiting a $\delta$-function in
$df/d\Theta$); a mock Jupiter model where the power-law index varies
from $1$ in the interior to $1.8$ in the envelope, with the transition
occurring at $r/R=0.98$ (solid lines, having a jump in $df/d\Theta$)
and spanning a range of $\Delta r/R \sim 0.002$ (FWHM of the spike in
$d^2f/d\Theta^2$); a similar model but with the transition occurring
over a range of $\Delta r \sim 0.02$ (dashed curves, the one with
smooth $df/d\Theta$). Analytically, we expect scalings of $1/n$,
$1/n^2$ switching to $1/n^{3.6}$ when $n > 500$, and $1/n^2$ switching
to $1/n^{3.6}$ when $n > 50$, for the three models,
respectively. These scalings are marked here as the three dot-dashed
lines.}
\label{fig:newpowerlaw} 
\end{figure}

\begin{figure}
\centerline{\psfig{figure=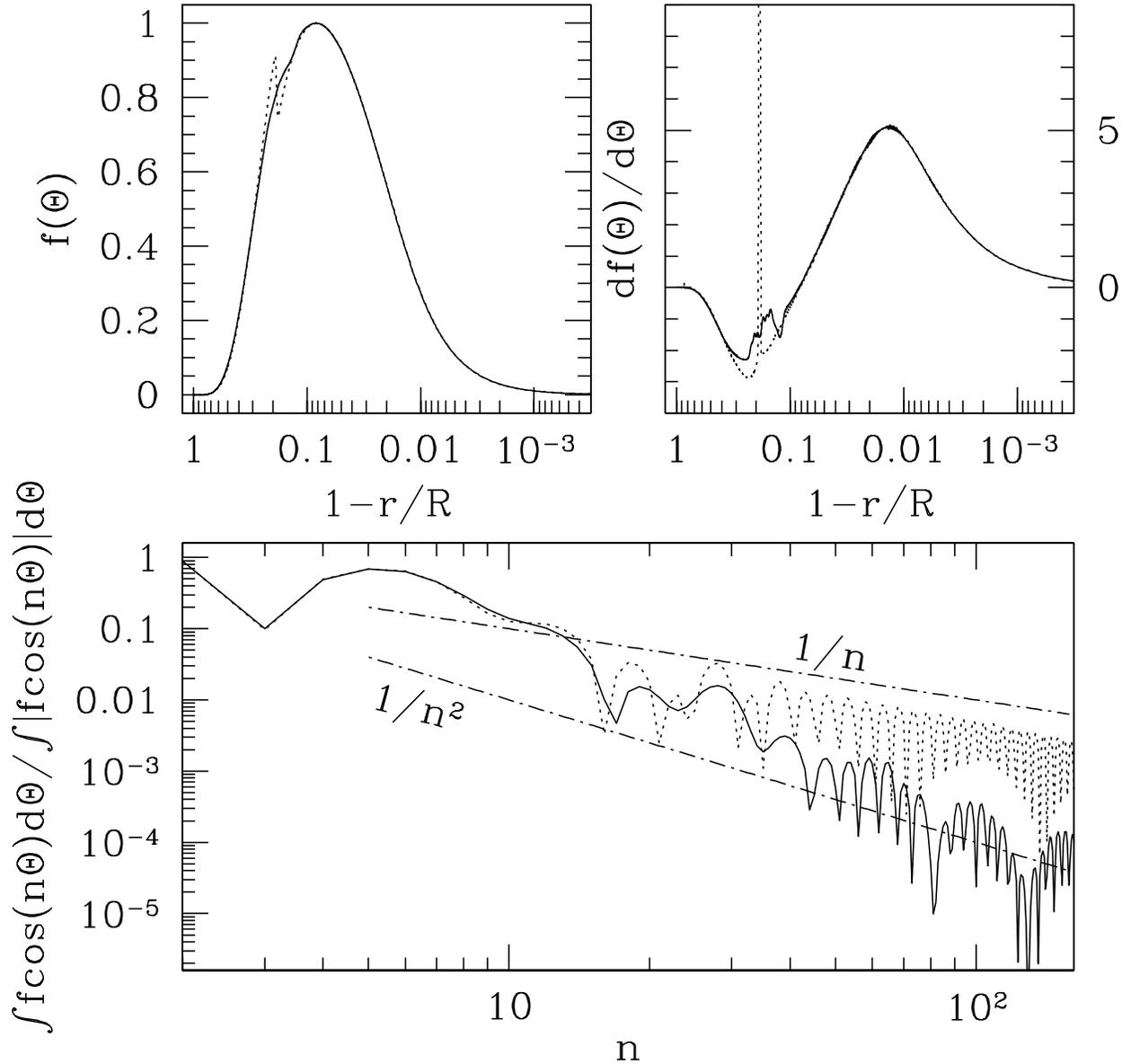,width=1.00\hsize}}
\caption{
Same as Fig. \ref{fig:newpowerlaw} but with the density profile taken
from two realistic Jupiter models: models B \& D as in
\citet{guillotreview}.
Model B (solid curves) is based on an interpolated equation of state
with no core and no density discontinuity across the metallic hydrogen
phase transition region at $r/R\sim 0.8$ -- but the first derivative
of density is discontinuous there ($df/d\Theta$ jumps by $\sim 50\%$).
Overlap integral in model B is expected to suffer a cancellation with
a $1/n^2$ scaling (lower panel). 
The sharp transition in the equation of state around $r/R \sim 0.98$,
with a FWHM for $d^2 f/d\Theta^2$ of $\Delta r/R \sim 0.02$, also
contributes to this scaling. But this contribution falls off sharply
for $n \gg 1/0.02 \sim 50$.
%
Model D (dotted curves) has a $10 M_\earth$ solid core, and is based
on PPT equation of state with the phase transition being first-order,
giving rise to a fractional density jump of $\sim 20\%$. This is
seen here as the jump in $f(\Theta)$ and the spike in $df/d\Theta$.
Overlap integral in model D is dominated by the density jump and it
scales roughly as $1/n$, as expected.
These results are insensitive to core sizes, since the $r^6$ scaling
in $f(\Theta)$ near the center suppresses any influence from the inner
boundary condition.
Moreover, severity of cancellation calculated for actual inertial-mode
eigenfunctions is expected to be one power of $n$ steeper than those
presented here, due to cancellation in the angular direction.
}
\label{fig:newpowerlaw-j} 
\end{figure}

We show that when a density discontinuity is superimposed to a single
power-law model (dotted curves in Fig. \ref{fig:newpowerlaw}), the
overlap integral indeed scales as $1/n$. Also, if the model has a
sharp (but continuous) transition in the value of $\beta$ over a
radius of $\Delta r$ (dashed and solid curves in
Fig. \ref{fig:newpowerlaw}), the integral scales as $1/n^2$ for $n
\leq \Delta r/R$, while for higher $n$ values, it behaves as is predicted by 
equation \refnew{eq:evenodd2}. We have also studied integration
results for two realistic Jupiter models taken from
\citet{guillotreview} (models B \& D). Model D has a first-order phase
transition (dotted curve in Fig. \ref{fig:newpowerlaw-j}) and so its
overlap integral scales as $1/n$; while the same phase transition is
considered to be second-order in model D, and the resulting
discontinuity in density gradient (as well as the equation of state
transition at $r/R \sim 0.98$, see Appendix \ref{subsec:density})
causes the integral to scale as $1/n^2$.


When the density profile is not a single power-law (as is the case
in this section), we could not solve for inertial-mode eigenfunctions
exactly. We could only obtain an approximate solution that is good to
the second order in wavenumber (${\cal O}(\lambda^2)$, see Paper
I). It is reasonable to suspect that the overlap results obtained by
integrating such an approximate solution deviate from the true one.  A
definite answer to this suspicion will likely be provided by full
numerical solution.  However, we argue below that the deviation should
be unimportant. 

The result of integrating a fast oscillation function, as is shown in
this section and Appendix \ref{sec:polyn}, depends only on the
boundary behavior and interior discontinuities in the envelope of such
a function. It does not depend on the exact shape of the function in
the interior. Our approximate solution to the inertial-modes is exact
near the surface, and is sufficiently accurate near the center (where
the WKB approximation works well). Moreover, when a density
discontinuity (or discontinuity in density derivatives) is present
inside the WKB region, as inertial-modes are insensitive to density
structure, the solution is not expected to deviate qualitatively from
the approximate solution that does not take this into account.

In confirming the scalings derived in this section, we have only
integrated the toy-model ($f(\Theta \cos n\Theta$), instead of
integrating inertial-mode eigen-functions (in
Fig. \ref{fig:singlebeta} we integrate both). One can similarly argue
that integrating an appropriately chosen $f(\Theta)$ is equivalent of
integrating the real function. In fact, our toy model should produce
results both qualitatively and quantitatively similar to that obtained
using the actual eigenfunctions, one can almost make do without
detailed knowledge of the latter.es such a density discontinuity.

Lastly, independent of the radial profile, integration in the angular
direction always introduces a factor of $1/n$ cancellation.

\end{appendix}

\end{document}